\documentclass[twocolumn,tighten,times]{aastex63}

\usepackage{tabularx}
\usepackage{natbib}
\usepackage{amsmath}
\usepackage{amssymb}
\usepackage{graphicx}
\usepackage{subfigure}
\usepackage{xcolor}
\usepackage{epsfig}
\usepackage{apjfonts}
\usepackage{soul}

\bibpunct[,]{(}{)}{;}{a}{}{,}
\allowdisplaybreaks[1]

\newcommand{\Msun}{\ifmmode {M_{\odot}}\else${M_{\odot}}$\fi} 

\newcommand{\sign}[0]{{\rm sgn}}

\shorttitle{Fast ejecta from NS mergers}
\shortauthors{Dean, Fern\'andez, \& Metzger}

\begin{document}

\title{Resolving the fastest ejecta from binary Neutron Star mergers: implications for electromagnetic counterparts}

\author[0000-0001-9364-4785]{Coleman Dean} 
\affil{Department of Physics, University of Alberta, Edmonton, AB T6G 2E1, Canada.}

\author[0000-0003-4619-339X]{Rodrigo Fern\'andez}
\affil{Department of Physics, University of Alberta, Edmonton, AB T6G 2E1, Canada.}

\author[0000-0002-4670-7509]{Brian D. Metzger}
\affil{Department of Physics and Columbia Astrophysics Laboratory, Columbia University, Pupin Hall, New York, NY 10027, USA}
\affil{Center for Computational Astrophysics, Flatiron Institute, 162 5th Ave, New York, NY 10010, USA} 

\begin{abstract}
We examine the effect of spatial resolution on initial mass ejection in
grid-based hydrodynamic simulations of binary neutron star mergers.  The subset
of the dynamical ejecta with velocities greater than $\sim 0.6$\,c can generate
an ultraviolet precursor to the kilonova on $\sim$\,hr timescales and
contribute to a years-long non-thermal afterglow.  Previous work has found
differing amounts of this fast ejecta, by one- to two orders of magnitude, when
using particle-based or grid-based hydrodynamic methods.  Here we carry out a
numerical experiment that models the merger as an axisymmetric collision in a
co-rotating frame, accounting for Newtonian self-gravity, inertial forces, and
gravitational wave losses. The lower computational cost allows us to reach
spatial resolutions as high as $4$\,m, or $\sim 3\times 10^{-4}$ of the stellar
radius.  We find that fast ejecta production converges {to within $10\%$} 
for a cell size of {$20$\,m.} This suggests that 
fast ejecta quantities found in existing grid-based
merger simulations are unlikely to increase to the level needed to match
particle-based results upon further resolution increases. The resulting
neutron-powered precursors are in principle detectable out to distances
$\lesssim 200$\,Mpc with upcoming facilities.  
We also find that head-on collisions at the free-fall speed,
relevant for eccentric mergers, yield fast and slow ejecta quantities of order
$10^{-2}M_\odot$, with a kilonova signature distinct from that of
quasi-circular mergers.
\end{abstract}


\section{Introduction}
\label{sec:intro}

Detection of electromagnetic (EM) emission from neutron star (NS) mergers
provides additional information beyond that contained in the gravitational wave
(GW) signal, as was demonstrated for GW170817
\citep{ligo_gw170817_multi-messenger}.  This information allows probing the
merger environment (e.g., \citealt{levan_2017,blanchard_2017}), their use as
standard sirens for cosmology (e.g., \citealt{ligo_gw170817_H0}), constraining
their status as progenitors of short gamma-ray bursts 
(e.g., \citealt{ligo_gw170817_grb}), or assessing their contribution to the cosmic
production of $r$-process elements (e.g., \citealt{cowperthwaite_2017,
drout_2017,tanaka_2017,tanvir_2017}). Nucleosynthesis information is encoded in
the kilonova signal, which arises from material ejected at subrelativistic
speeds that is radioactively heated by freshily formed elements
\citep{li_1998,metzger_2010}.

The bulk of mass ejection in binary NS (BNS) mergers occurs in two ways.
First, material is ejected on the dynamical time from the collision interface
between the two stars or by tidal processes as the
stars merge 
(e.g., \citealt{hotokezaka_2013}).
Second, the accretion disk that forms after the merger ejects mass over longer
timescales (see, e.g., \citealt{FM16} for a review). In both cases, the
majority of the material is initially neutron-rich and moves at speeds
$\lesssim 0.3$\,c, which allows the $r$-process to proceed to completion, with
a composition pattern that depends on the  level of reprocessing by neutrinos (e.g.,
\citealt{wanajo2014,just_2015,martin2015,wu2016, roberts2017,lippuner_2017}).
This results in a kilonova signal that peaks in the optical or infrared band
and which evolves on a day to week timescale
\citep{kasen_2013,tanaka2013,barnes_2013,fontes2015}.

If a fraction of the ejected material expands on sufficiently short timescales, 
a freezout of the $r$-process can occur, with the ejecta consisting
primarily of free, unprocessed neutrons that eventually undergo beta decay
\citep{goriely_2014}. Freezout of neutrons requires the density to drop to 
$\sim 4\times 10^5$\,g\,cm$^{-3}$ on timescales shorter than $\sim 5$\,ms,
which maps well to ejecta with velocities $\gtrsim 0.6$\,c \citep{metzger_2015}.
Such ejecta can also generate EM emission
\citep{kulkarni_2005}, and if launched ahead of slower material, can provide an
ultraviolet \emph{precursor} to the kilonova that evolves on a timescale of
hours after the merger \citep{metzger_2015}. Detection of EM emission on a
timescale of hours could have differentiated among various models that account for
the kilonova from GW170817, but which diverge before the earliest EM observation
at $\sim 11$ hours post-merger \citep{arcavi_2018}. 

The existence of sufficient ejecta with the required speed to generate a
detectable neutron-powered precursor is not clear, however. A fast component with mass
$\sim 10^{-4}\,\Msun$ was first obtained in smoothed particle hydrodynamic
(SPH) simulations of BNS mergers \citep{bauswein_2013}, but grid-based
hydrodynamic simulations have found much smaller quantities
($\sim 10^{-7} - 10^{-5} \Msun$, e.g., \citealt{ishii_2018},\citealt{radice_2018b}),
making a potential kilonova precursor much harder to detect given the expected distance
to most sources ($\gtrsim 100$\,Mpc) and current EM sensitivity limits.  

More broadly, fast ejecta contributes to the non-thermal afterglow generated
by outgoing mass interacting with the interstellar medium \citep{nakar_2011}.
In particular, fast ejecta has been proposed as a possible origin for the X-ray excess
detected from GW170817 3 years after the merger
\citep{hajela_2021,nedora_2021}. Small quantities ($\sim
10^{-8}-10^{-7}M_\odot$) of fast ejecta have also been proposed to account for
the overall properties of the prompt gamma-ray burst emission from GW170817 via
breakout of a jet from a rapidly-expanding cloud \citep{beloborodov_2020}.  Finally,
fast ejecta can also be produced in eccentric mergers, which can produce nearly head-on
collisions at much higher radial velocities than in quasi-circular
mergers (e.g., \citealt{gold_2012,chaurasia_2018,papenfort_2018}).

The reliability of ejecta masses from grid-based simulations is tied to
how well the collision is spatially resolved, however. The spatial resolution of 
these simulations is usually limited by computational resources, with 
the finest grid spacings used to date being $63-86 \rm
m$ \citep{kiuchi_2017}. Properly resolving the surface layers of the star
requires grid sizes $< 10$\,m \citep{kyutoku_2014}.

Here we perform a numerical experiment to assess the spatial resolution dependence 
of fast ejecta from the collision interface of BNS mergers 
in grid-based hydrodynamic simulations.  To decrease the
computational cost, we solve the Newtonian hydrodynamics equations in
two-dimensional (2D) cylindrical symmetry with self gravity in a co-rotating frame,
to account for orbital motion, and with an approximate prescription for orbital
decay due to gravitational waves. These approximations, while losing accuracy
relative to a full three-dimensional (3D) setting, preserve the qualitative feature
of two sharp stellar edges colliding under the relevant force
environment, and allow us to reach grid sizes as low as $4$\,m ($\sim 3\times 10^{-4}$
of the NS radius).

The structure of the paper is the following. Section~\ref{sec:methods}
describes our physical assumptions, computational method, and choice of models.
Our results are presented in Section~\ref{sec:results}, with a general
overview of our baseline model, parameter dependencies, and comparison
with previous work. The observational implications of our results are presented in
Section~\ref{sec:em_implications}, and a summary and discussion follows 
in Section~\ref{sec:summary}.

\section{Methods}
\label{sec:methods}

\subsection{Physical Model and Approximations}
\label{sec:physical_model}

Our goal is to study the ejection of material during the initial collision
between the merging NSs and its immediate aftermath,
with a focus on the high-velocity tail of the ejecta velocity distribution.
Our numerical experiment attempts to capture the key features of the stellar 
collision in 2D, which allows for a much higher spatial resolution than is 
achievable in a full 3D configuration.

The hydrodynamic interaction between the two stars is influenced primarily by
three effects: gravity, orbital motion, and orbital decay due to
gravitational wave emission. Correspondingly, we neglect neutrino processes,
magnetic fields, and stellar rotation in our study. While these effects
can certainly influence mass ejection, they either provide sub-dominant corrections to
the main processes considered here or, in the case of magnetic fields, introduce
complications for comparing with previous work. Additionally, we do not keep track of 
the ejecta composition, and assess the feasibility of $r$-process freezout
based on ejecta velocity, which is a good proxy for expansion time \citep{metzger_2015}.

\begin{figure}
\centering 
\includegraphics*[width=\columnwidth]{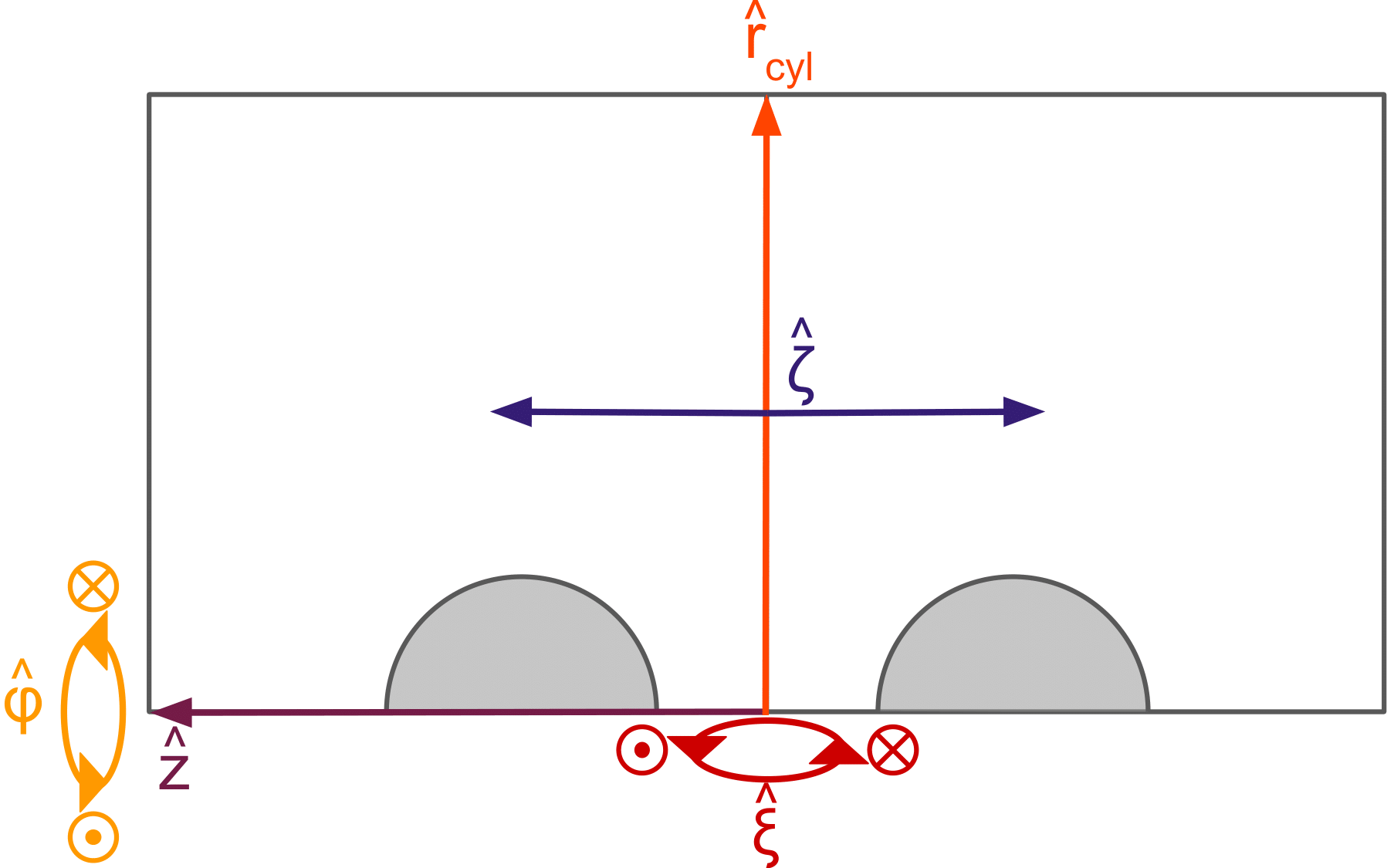} 
\caption{Schematic of the coordinate system used. An axisymmetric domain in 
cylindrical coordinates $(r_{\rm cyl},\varphi,z)$ is rotated $90^\circ$ counterclockwise
and the stars are placed along the symmetry axis ($z$). The orbital
angular momentum vector points in the cylindrical radial ($r_{\rm cyl}$) direction. Orbital 
motion occurs along the pseudo-azimuthal vector $\hat \xi = -{\rm sgn}(z)\hat\varphi$
and the distance from the rotation axis is described by the
auxiliary vector $\hat\zeta = {\rm sgn}(z)\hat z$.}
\label{fig:unit_vectors}
\end{figure}

We model the NS binary geometry in 2D by adopting an axisymmetric cylindrical coordinate
system, positioning the stars along the symmetry ($\hat z$) axis, thereby
accounting for their spherical geometry along the azimuthal ($\hat\varphi$) direction.
We then set the orbital angular momentum vector aligned with the cylindrical 
radial ($\hat r_{\rm cyl}$) axis, with orbital motion occurring along a 
pseudo-azimuthal vector  $\hat\xi = -\sign(z)\hat\varphi$ around this axis 
(Figure~\ref{fig:unit_vectors}). 

Orbital motion is quantified with a specific angular momentum scalar 
$j = \omega z^2$, where $\omega(r_{\rm cyl},z,t)$ is the orbital 
angular velocity ($t$ is time). We work in a co-rotating coordinate system 
around $\hat r_{\rm cyl}$ with constant angular frequency $\omega_0=2\pi/t_{\rm dyn}$, where
\begin{equation}
\label{eqn:tdyn_binary}
t_{\rm dyn} = \left[\frac{4\pi^2}{G(M_1+M_2)}d_0^3\right]^{1/2}
\end{equation}
is the initial (Newtonian) orbital period of the system, with $d_0$ the initial separation between the
stellar centers, and $M_{1,2}$ the stellar masses.

The centrifugal acceleration in this co-rotating frame is then
\begin{equation}
\label{eqn:centrifugal_force}
\mathbf{f}_{\rm cen} = \omega_0^2 |z|\, \hat\zeta = \omega_0^2 |z|\sign(z)\,\hat z  = z\omega_0^2\, \hat z, 
\end{equation}
where $\hat \zeta = \sign(z)\,\hat z$ is a unit vector that points away from the orbital
axis (Figure~\ref{fig:unit_vectors}). The Coriolis acceleration is given by
\begin{equation}
\label{eqn:coriolis_force}
\mathbf{f}_{\rm cor} = 2\omega_0 \left[v_\xi\hat\zeta - v_\zeta\hat\xi \right] =
2\omega_0 \sign(z)\left[v_\xi\hat{z} -v_z\hat\xi\right],
\end{equation}
where
\begin{equation}
v_\xi = \frac{j}{|z|} - |z|\omega_0
\end{equation}
is the co-rotating frame velocity along $\hat\xi$, $v_\zeta = \sign(z)v_z$ is the speed
away from the orbital angular momentum axis, and $v_z$ is the velocity along $\hat z$.
With this formulation, matter moving toward the axis ($v_\zeta < 0$)
is accelerated in the $+\hat\xi$ direction, while matter rotating faster
than the coordinate system ($v_\xi >0$) is pushed away from the rotation
axis ($+\hat\zeta$), thus following the standard behavior of the Coriolis
force. The azimuthal term in the Coriolis acceleration acts as a source
term for the specific angular momentum
\begin{equation}
\label{eqn:dldt_coriolis}
\left(\frac{\partial j}{\partial t} \right)_{\rm cor} = (\mathbf{r}\times \mathbf{f}_{\rm
cor})_{r_{\rm cyl}} = -2\omega_0 z v_z.
\end{equation}
We neglect other components of the torque given the symmetry of our coordinate
system (the $\zeta$ component of the Coriolis acceleration adds or subtracts from 
the centrifugal acceleration). 

\begin{figure}
\centering 
\includegraphics*[width=\columnwidth]{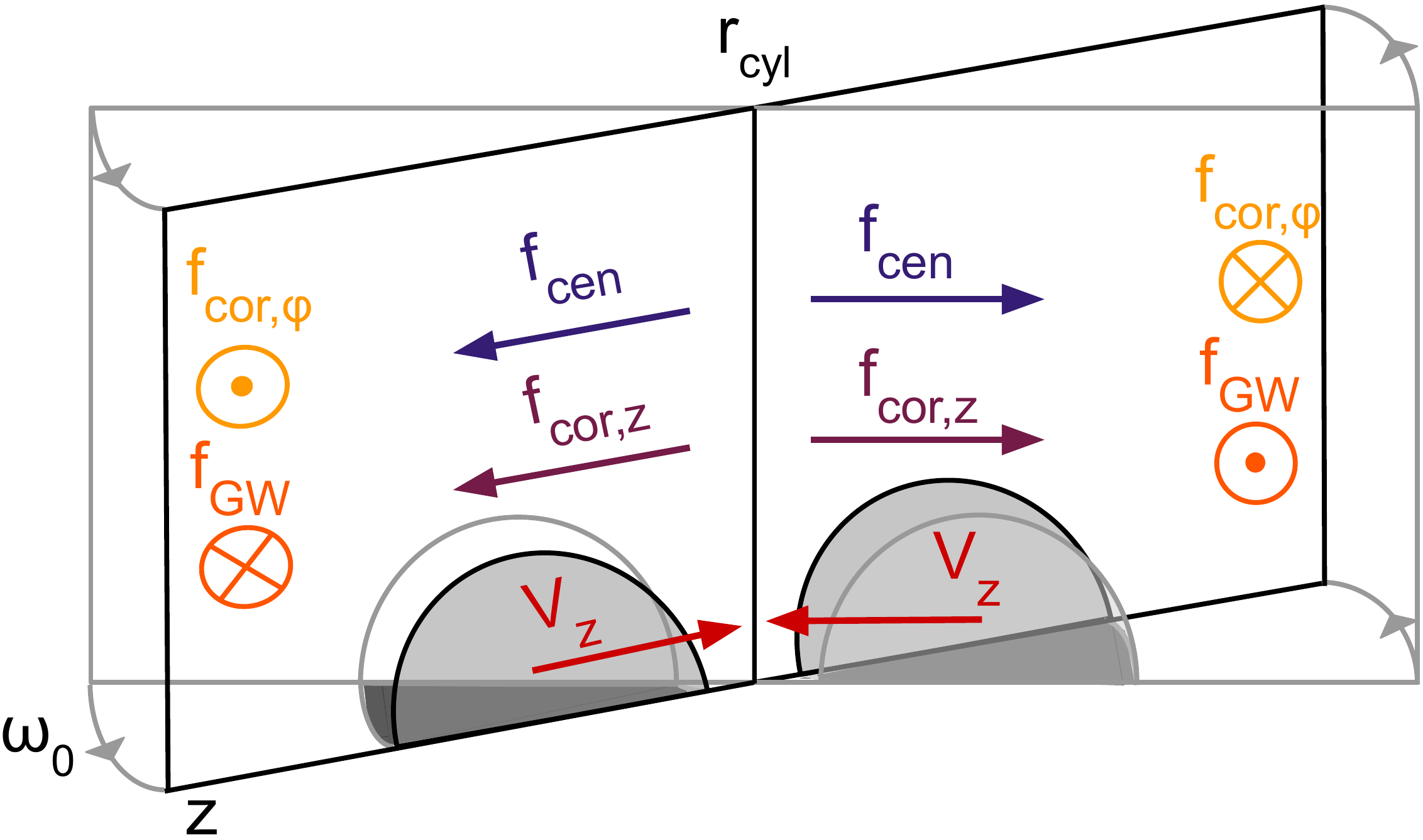} 
\caption{Schematic of the numerical experiment. A corotating frame is employed, with constant angular velocity $\omega_0$ set 
to the initial Keplerian value. Matter thus
experiences centrifugal and Coriolis accelerations 
(equations~\ref{eqn:centrifugal_force}-\ref{eqn:coriolis_force}).
Orbital motion is quantified with a space- and time-dependent specific angular momentum scalar 
$j =\omega z^2$, which 
can be modified by the component of the
Coriolis acceleration in the orbital direction (equation~\ref{eqn:dldt_coriolis})
and by gravitational wave emission (equation~\ref{eqn:dldt}). 
The stars with masses $M_1$ and $M_2$ are initialized at an initial
separation $d_0$ measured from their centers, 
with the center of mass at the origin, and with 
velocities in the $z$-direction toward the rotation axis,
resulting in the 
acceleration directions shown.
The intrinsic symmetry of the cylindrical coordinate
system accounts for the sphericity of the stars.}
\label{fig:fictitious_force}
\end{figure}

The correctness of this formulation is verified 
by the maintenance of a stable Keplerian orbit in the absence of gravitational wave 
losses (Section \ref{sec:hydrodynamics}). A schematic of the relative direction of
the inertial accelerations is shown in Figure~\ref{fig:fictitious_force}.

\subsection{Numerical Hydrodynamics}
\label{sec:hydrodynamics}

We use {\tt FLASH} version 4.5 \citep{fryxell00,dubey2009} to solve the
equations of Newtonian hydrodynamics in 2D cylindrical coordinates in a 
co-rotating frame, with source terms due to self-gravity and gravitational
wave losses:
\begin{eqnarray}
\label{eqn:mass_conservation}
\frac{\partial \rho}{\partial t} + \nabla\cdot (\rho \mathbf{v})  & = &  0 \\
\label{eqn:momentum_conservation}
\frac{D\mathbf{v}}{Dt}  & = & -\frac{\nabla P}{\rho} -\nabla\Phi + \mathbf{f}_{\rm cen} + (\mathbf{f}_{\rm cor}\cdot\hat\zeta)\hat\zeta \\
\label{eqn:angular_mom_conservation}
\frac{Dj}{Dt} & = & \left(\frac{\partial j}{\partial t} \right)_{\rm gw} + \left(\frac{\partial j}{\partial t} \right)_{\rm cor}\\
\label{eqn:energy_conservation}
\frac{D\epsilon}{Dt}  - \frac{P}{\rho^2}\frac{D\rho}{Dt}
              & = & 0\\
\label{eqn:poisson}
\nabla^2 \Phi  & =  & 4 \pi G \rho \\
\end{eqnarray}
where $D/Dt \equiv (\partial/\partial t + \mathbf{v}\cdot\nabla)$, 
$\mathbf{v} = v_{\rm rcyl}\hat r_{\rm cyl} + v_z \hat z$ is the poloidal
velocity,
$\rho$ is the density, $P$ is the total gas pressure, $\epsilon$ is the total specific
internal energy, $\Phi$ is the gravitational potential, and $G$ is the gravitational constant.
The system of equations is solved with the dimensionally-split Piecewise-Parabolic Method
(PPM; \citealt{colella84}), the multipole self-gravity solver of
\cite{couch_2013}, and is closed with a piecewise polytropic equation of state
(EOS).

The EOS contains a cold component with 4-segments
\begin{equation}
\label{eqn:pwp}
  P_{{\rm c},i} = K_i \rho^{\Gamma_i}\qquad i = \{0,1,2,3\} ,
\end{equation}
where the adiabatic indices $\Gamma_i$, transition densities $\rho_i$, 
and transition pressure $P_{\rm c,1}$ are taken from \citet{read_2009}. 
This cold component connects continuously at low density to a \texttt{SLy} EOS for the
crust \citep{douchin_2001}. The EOS also includes a thermal component (e.g.,
\citealt{bauswein_2010}), such that the total pressure satisfies

\begin{eqnarray}
P              & = & P_{\rm c} + P_{\rm th}\\
P_{\rm th}     & = & (\Gamma_{\rm th} - 1)\rho \epsilon_{\rm th}\\
\epsilon_{\rm th} & = & \epsilon - \epsilon_{\rm c}.
\end{eqnarray}
where the subscripts ``c" and ``th" denote the cold and thermal components,
respectively. We adopt $\Gamma_{\rm th} = 5/3 \simeq 1.67$ as an intermediate
value in the interval $[1.5,2]$, which was found by \citet{bauswein_2010} to
bracket the behavior of micro-physical, finite-temperature EOS's.

We include the effect of gravitational wave losses on the orbit through 
a source term that modifies the specific angular momentum
\begin{equation}
\label{eqn:dldt}
\left(\frac{\partial j}{\partial t} \right)_{\rm gw} =
\frac{1}{2}\frac{j}{d}\left(\frac{\partial d}{\partial t}\right)_{\rm gw}
\end{equation}
where we assume a Keplerian dependence on the instantaneous orbital separation $d$
between the centers of mass of each star (semimajor axis of the reduced mass), 
$j\propto d^{1/2}$. The rate of change of $d$ is taken to be the standard expression 
for two point masses with $e=0$ \citep{peters_1964},
\begin{equation}
\label{eqn:adot_gw}
\left(\frac{\partial d}{\partial t}\right)_{\rm gw} = -\frac{64}{5} \frac{G^3}{c^5}
\frac{(M_1+M_2)M_1M_2}{d^3}.
\end{equation}
While these equations are strictly valid only for point masses,
we apply this source term to all stellar material that has non-zero $j$ 
(i.e., material ejected after the collision). Once $j$ reaches zero,
the source term is also set to zero. For equal-mass binaries, we set 
$d = 2|z|$ in equations~(\ref{eqn:dldt}) and (\ref{eqn:adot_gw}).
For asymmetric binaries, we need to account for the mass ratio $q=M_2/M_1$, and hence set
$d=(1 +1/q)|z_1|$ or $d=(1+q)|z_2|$ for all matter on the side of star 1 or star 2 relative
to $z=0$, respectively.  Aside from this substitution, the mass ratio dependence scales out 
of equation~(\ref{eqn:dldt}).

\begin{figure}
\centering
\includegraphics*[width=\columnwidth]{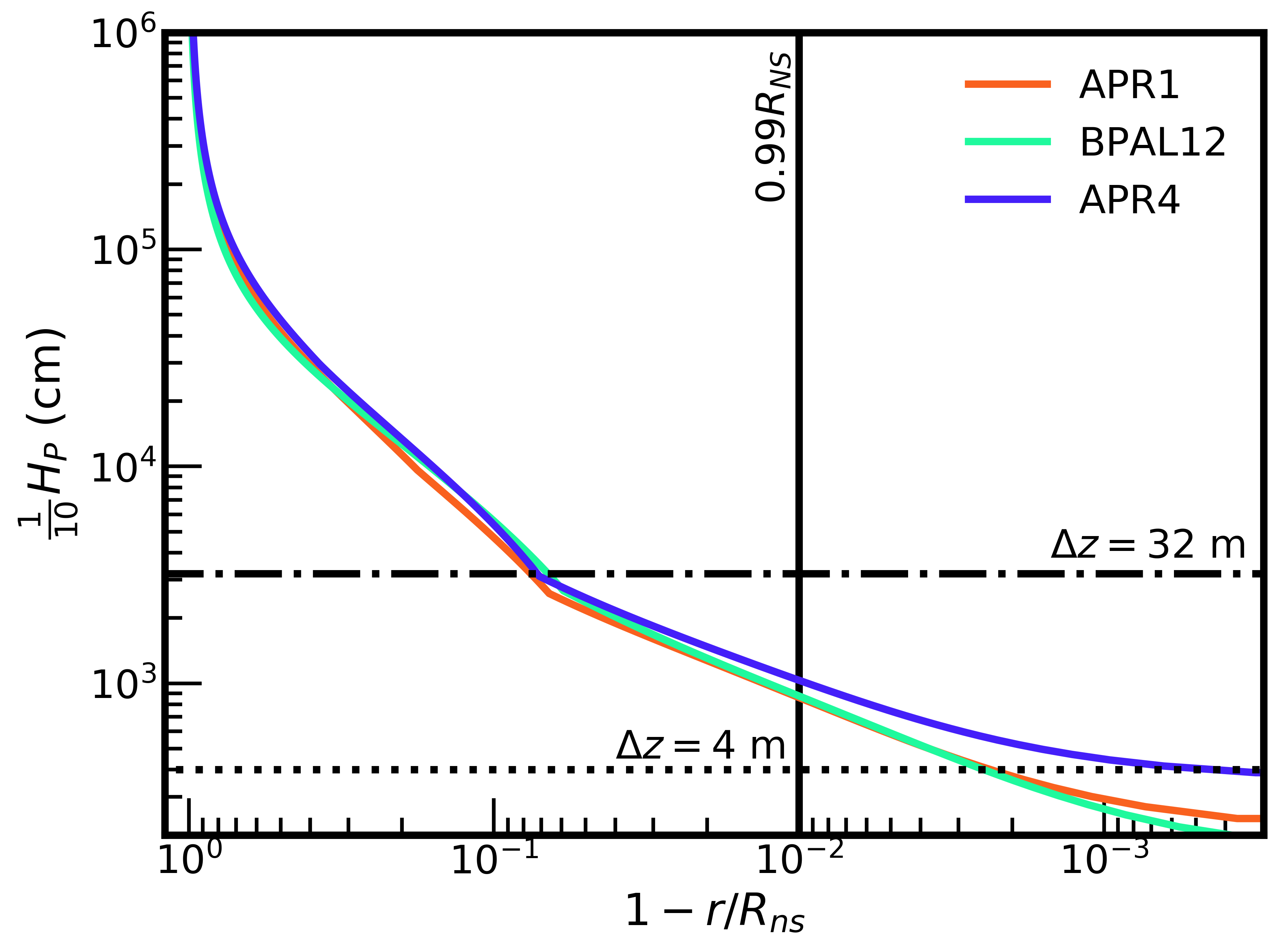}
\caption{Spatial resolution 
in terms of relevant length scales in the 
problem. Solid curves show $1/10$th of the pressure scale height $H_{\rm p}$
as a function of exterior radius in a $1.4M_\odot$ NS constructed with 
the \texttt{APR4} EOS (purple), \texttt{BPAL12} EOS (teal), and 
\texttt{APR1} EOS (orange). Also shown as horizontal lines are our 
baseline resolution $\Delta r_{\rm cyl}=\Delta z=32$\,m (dot-dashed)
and the finest resolution used, $\Delta r_{\rm cyl}=\Delta z=4$\,m (dotted).
The latter covers $H_p$ with $10$ cells out to $>99.9\%$ of the stellar radius, 
corresponding to an exterior mass of $<9\times10^{-4}\Msun$ for the \texttt{APR4} EOS.}
\label{fig:pressure_scale_height}
\end{figure}

The computational domain spans the range $[0,90]$\,km in $r_{\rm cyl}$ and $[-90,90]$\,km in $z$, and 
is discretized with a uniform grid with square cells $\Delta r_{\rm cyl} = \Delta z$. We choose our
resolution in relation to the pressure scale height $H_{\rm p} = P/(\rho
|\nabla \Phi|)$ near the stellar surface (Figure~\ref{fig:pressure_scale_height}).
Given the numerical dissipation properties of PPM (e.g.,
\citealt{porter_1994}), we consider a length scale as resolved if we can cover
it with 10 computational cells. Our baseline grid spacing is $\Delta r_{\rm cyl}=\Delta z = 32$\,m  
in all models (Section \ref{sec:initial_conditions}), 
which resolves the pressure scale height out to $\sim 93\%$ of the stellar radius. 
Our finest resolution, $\Delta r_{\rm cyl}=\Delta z = 4$\,m,
reaches beyond 99.9\% of the stellar radius $R_{\rm ns}$,
enclosing all but the outermost $\sim 9 \times
10^{-4}M_{\odot}$ of the stellar mass. 

The boundary conditions are reflecting at $r_{\rm cyl}=0$, and outflow
at all other domain limits.
The orbital configuration is tested by 
initializing the stars in a Keplerian orbit, with an
initial separation $d_0 \simeq 8R_{\rm ns}$,
and verifying that the stars maitain their initial positions in the absence of gravitational wave losses,
with only minor in-place oscillations. Based on this stationarity test we
employ $128$ multipoles for self-gravity in all of our runs. 

\subsection{Initial Conditions}
\label{sec:initial_conditions}

\begin{figure}
\includegraphics*[width=\columnwidth]{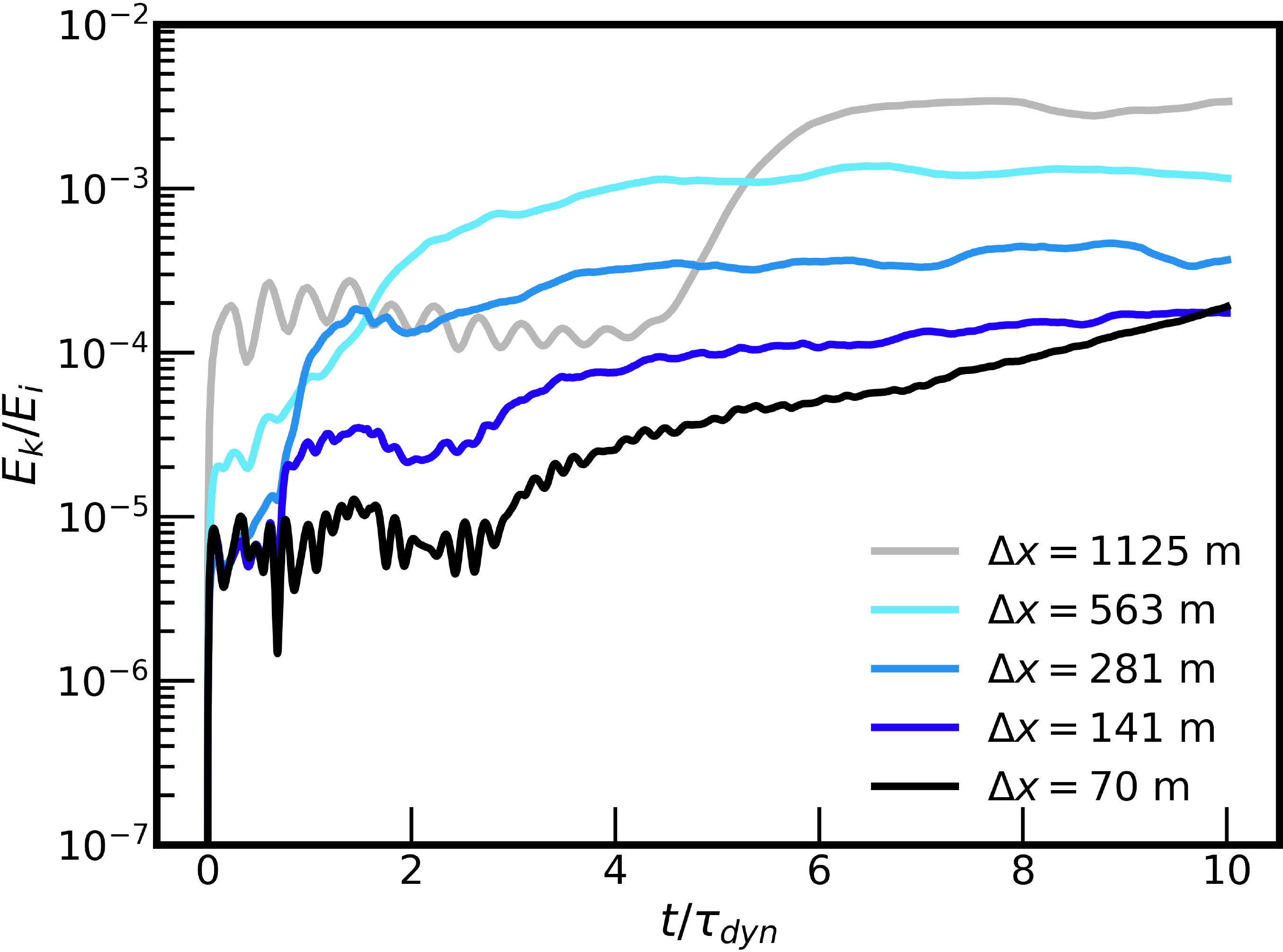}
\caption{Hydrostatic equilibrium test of an isolated $1.4M_\odot$ NS constructed
with the APR4 EOS at the center of the computational domain, showing the
ratio of kinetic energy $E_{\rm k}$ to internal energy $E_{\rm i}$ 
in stellar material as a function of time, in units of the dynamical time of 
the star $\tau_{\rm dyn}\simeq 0.07$\,ms.  
Curves correspond to different spatial resolutions, as labeled.}
\label{fig:E_ratio}
\end{figure}

Neutron stars are constructed by solving the Newtonian hydrostatic equilibrium equations 
using the EOSs described in Section \ref{sec:hydrodynamics}. We test the solution by 
evolving an isolated star centered at the origin for $10$ stellar dynamical 
times $\tau^{\rm ns}_{\rm dyn} \simeq (2\pi)^{-1} (R_{\rm NS}/d_0)^{3/2}\,t_{\rm dyn}$, 
and verifying that it remains close to a steady-state with low kinetic energy (Figure~\ref{fig:E_ratio}).

Simulations are initialized with 
the two stellar centers placed on the $z$-axis at a
separation $d_0$ slightly larger 
than the sum of the stellar radii, and located such
that the center of mass of the system is at the origin, 
$|z_1/z_2| = M_2 / M_1 = q$, with $|z_1|$ and $|z_2|$
the initial $z$ coordinates of the corresponding stellar centers

In our default configuration, the interior of each star is
assigned the Keplerian velocity of the center of mass uniformly,
with a correction for
the inspiral of the orbit\footnote{This correction
is obtained by assuming energy conservation in an orbit decaying by gravitational waves, and
is negligible for $d \gtrsim 2R_{\rm NS}$. It is nevertheless included for completeness.},
thus initially $v_\xi \simeq 0$ at the center of mass of each star.
The initial velocity along the $z$ axis is set to the value implied by the
decay rate of the orbital separation due to gravitational wave emission 
(equation~\ref{eqn:adot_gw}), with a characteristic magnitude $\sim 2\times 10^9$\,cm\,s$^{-1}$. 
To probe the sensitivity of our results to the initial conditions, we also evolve a head-on collision 
model that removes rotation, intertial forces, and gravitational wave losses, as well as one
that sets the velocity along the $z$ axis to the free-fall value.

The stars are initially embedded in an ambient medium of mass $\sim2.3\times10^{-8}\Msun$,
density $10^4~\rm{g}/\rm{cm}^3$, and constant pressure
$10^{25}~\rm{dyn}/\rm{cm}^2$, with the remaining thermodynamic variables
determined by the EOS. The ambient mass is negligible relative to characteristic
ejecta masses of interest, and hence it should not significantly influence the velocity of this ejecta.

\begin{table*}
\centering
\caption{Models evolved and results.}
\begin{tabular}{ccccccccccccc}
\hline
Model & EOS  & Inertial Force &$M_1$ & $M_2$ &
$R_{\rm 1.4}$ & $\Delta x$ & $M_{\rm ej}$ & \multicolumn{3}{c}{$M_{v\ge 0.6 c}$}
& \multicolumn{2}{c}{$E_{k}$}\\ 
 & & Treatment & & & & & & total & contact & orbital & $>0.3c$ & $>0.6c$ \\
 & & &($\Msun$) & ($\Msun$) & (km) & (m) & ($\Msun$) & & ($\Msun$) & & \multicolumn{2}{c}{($10^{49}$\,erg)} \\
\noalign{\smallskip}
\hline
base & APR4 & All forces & 1.4  & 1.4  & 12.6 & 32 & $2.5\times 10^{-5}$ &
$1.4\times 10^{-5}$ & $1.3\times 10^{-5}$ & $6.6\times 10^{-8}$ & $1.0$ & $0.83$ \\ 
\noalign{\smallskip}
R281* &  &  &      &      &  & 281.3 & $1.9\times 10^{-5}$ & $8.0\times 10^{-6}$ &
$8.0\times 10^{-6}$ & $6.4\times 10^{-8}$ & $0.57$ & $0.35$\\ 
R141* &  &  &  &  &  & 140.6 & $2.7\times 10^{-5}$ & $1.5\times 10^{-5}$ & $1.5\times
10^{-5}$ & $1.8\times 10^{-8}$ & $1.0$ & $0.79$ \\ 
R70*  &  &  &  &  &  & 70.3  & $2.8\times 10^{-5}$ & $1.5\times 10^{-5}$ & $1.5\times
10^{-5}$ & $6.3\times 10^{-8}$ & $1.2$ & $0.98$ \\ 
R16    &  &  &  &  &  & 16.0   & $2.2\times 10^{-5}$ & $1.1\times 10^{-5}$ &
$1.1\times 10^{-5}$ & $1.0\times 10^{-7}$ & $0.98$ & $0.78$ \\ 
R4    &  &  &  &  &  & 4.0   & $2.6\times 10^{-5}$ & $1.2\times 10^{-5}$ & $1.2\times
10^{-5}$ & $1.2\times 10^{-7}$ & $1.0$ & $0.77$ \\ 
\noalign{\smallskip}
APR1 & APR1 & & & & 11.0 & 32 & $5.4\times 10^{-5}$ & $2.6\times 10^{-5}$ & $2.6\times
10^{-5}$ & $5.4\times 10^{-7}$ & $2.6$ & $2.1$\\ 
BPAL12 & BPAL12 & & & & 14.1 & & $1.4\times 10^{-6}$ & $1.3\times 10^{-8}$ &
$1.3\times 10^{-8}$ & $1.1\times 10^{-10}$ & $0.02$ & $0.0006$\\ 
\noalign{\smallskip}
OR & APR4 & On rails & & & 12.6 & & $1.3\times10^{-2}$ & $5.7\times 10^{-3}$ & 
$2.2\times 10^{-3}$ & $3.6\times 10^{-3}$ & $430$ & $300$\\
OR70 & & & & & & 70.3 & $1.6\times10^{-2}$ & $6.4\times 10^{-3}$ & 
$1.9\times 10^{-3}$ & $4.5\times 10^{-3}$ & $520$ & $360$\\
OR141 & & & & & & 140.6 & $1.3\times10^{-2}$ & $7.0\times 10^{-3}$ & 
$1.7\times 10^{-3}$ & $5.3\times 10^{-3}$ & $460$ & $360$\\
FF & & Free-fall & & & & 32 & $3.6\times 10^{-2}$ & $1.2\times 10^{-2}$ & 
$1.5\times 10^{-4}$ & $1.2\times 10^{-2}$ & $950$ & $530$ \\ 
FF70 & & & & & & 70.3 & $3.2\times 10^{-2}$ & $9.1\times 10^{-3}$ & 
$2.0\times 10^{-4}$ & $8.9\times 10^{-3}$ & $830$ & $410$\\ 
FF141 & & & & & & 140.6 & $3.9\times 10^{-2}$ & $1.1\times 10^{-2}$ & 
$2.0\times 10^{-4}$ & $1.1\times 10^{-2}$ & $940$ & $510$\\ 
\noalign{\smallskip}
M1.2\_1.2 & & All forces  & 1.2 & 1.2 & & 32.0 & $1.9\times 10^{-5}$ & $8.2\times
10^{-6}$ & $8.2\times 10^{-6}$ & $2.5\times 10^{-8}$ & $0.74$ &
$0.53$ \\ 
M1.7\_1.7 & & & 1.7 & 1.7 & & & $5.0\times 10^{-5}$ & $2.5\times 10^{-5}$ &
$2.5\times 10^{-5}$ & $8.0\times 10^{-7}$ & $1.9$ & $1.5$ \\ 
\noalign{\smallskip}
M1.5\_1.3 & & & 1.5 & 1.3 & & & $2.7\times 10^{-4}$ & $2.7\times 10^{-5}$ &
$2.7\times 10^{-5}$ & $5.1\times 10^{-7}$ & $5.1$ & $1.2$ \\ 
M1.6\_1.2 & & & 1.6 & 1.2 & & & $2.5\times 10^{-4}$ & $1.4\times 10^{-5}$ &
$1.2\times 10^{-5}$ & $1.6\times 10^{-6}$ & $4.5$ & $0.61$ \\ 
\hline
\label{tab:models}
\end{tabular}
\begin{flushleft}
{\bf Note:} Columns from left to right show model name, EOS, inertial force
setup as defined in Section \ref{sec:initial_conditions}, NS masses, radius of a
$1.4 \Msun$ NS, cell size ($\Delta z = \Delta r_{\rm cyl}$), total unbound
mass ejected, fast ($v_r\ge 0.6$\,c) unbound ejecta, cumulative kinetic energy
of matter travelling at $v_r > 0.3$\,c and similarily for $v> 0.6$\,c. 
The latter is shown in the last 3 columns as total amount, and broken up by angular direction. 
Ejecta masses represent cumulative amounts
launched by the end of the default simulation time (1.6 ms).
Models marked with a star (low-resolution) were evolved for 10 times longer than
the rest of the set.
\end{flushleft}
\end{table*}

\subsection{Models Evolved}
\label{sec:models}

Table~\ref{tab:models} shows all of our models and their initial parameters. 
Our baseline case consists of two NSs with equal mass $M_1 = M_2 = 1.4M_\odot$ 
built with the \texttt{APR4} EOS \citep{akmal_1998}, yielding a radius $R_{\rm ns}=12.6$\,km.
This lies within the radius range allowed by GW170817 \citep{ligo_gw170817_radii},
thus yielding a realistic compactness. The default spatial resolution is 
$\Delta r_{\rm cyl} = \Delta z = 32$\,m, and the default evolution time is
$\simeq 1.2\, t_{\rm dyn}$ (equation~\ref{eqn:tdyn_binary}), corresponding to $1.6$\,ms.
This time interval is sufficient to achieve complete ejection of fast material
in our simulations.

To probe the effect of spatial resolution, the fiducial configuration is also 
evolved at grid spacings $\Delta z = \{281,141,70\}$\,m, which overlap with values
used in previous 3D numerical relativity simulations of BNS mergers. The
lower computational cost of these models allow us to evolve them
for $\simeq12\,t_{\rm dyn}$, or $16$\,ms. Two high-resolution models that employ
$\Delta z = \{16,4\}$\,m are evolved for $\simeq1.2t_{\rm dyn}$ to test for convergence.

We probe the effect of varying the EOSs -- and thus the compactness of the NSs -- with 
two models that use \texttt{APR1} \citep{akmal_1998} and \texttt{BPAL12} \citep{zuo_1999}, which yield neutron star radii
$11$\,km and $14.1$\,km for a stellar mass $1.4M_\odot$, respectively. All other simulation
parameters (aside from the initial separation, which depends on the stellar radius) are
identical to those in the default configuration. 

Likewise, we probe the effect of our force prescription by evolving two models
that remove rotation, inertial forces, and gravitational wave losses from the baseline
configuration. In one case (model \texttt{OR}, `on rails'), we leave the initial collision 
velocity along $z$ unchanged from the baseline configuration, corresponding to the rate of
decay of the orbital separation by gravitational waves (\S\ref{sec:hydrodynamics}). 
The other model (\texttt{FF}) sets the collision velocity to the free-fall speed.
Both of these models are evolved at the default resolution as well as at coarser grid
sizes $\Delta z = {141,70}$\,m, to probe the sensitivity to mass ejection to
this parameter (models OR70, OR141, FF70, and FF141).

The effect of changing the total binary mass at constant mass ratio is studied with 
models \texttt{M1.2\_1.2} and \texttt{M1.7\_1.7}, which set the total mass to $1.2+1.2\,M_\odot$
and $1.7+1.7M_\odot$, respectively. We ignore here the possibility of prompt collapse,
which is possible for the model with the highest total mass. Models 
M1.5\_1.3 and M1.6\_1.2 keep the total mass constant at $2.8\Msun$, but change the
mass ratio to $q=0.87$ ($1.5\Msun+1.3\Msun$) and $q=0.75$ ($1.6\Msun +
1.2\Msun$), respectively. The initial stellar positions and velocities of these
asymmetric cases are consistent with the description in 
Section~\ref{sec:initial_conditions}.

\section{Results}
\label{sec:results}

\begin{figure*}
\includegraphics*[width=\textwidth]{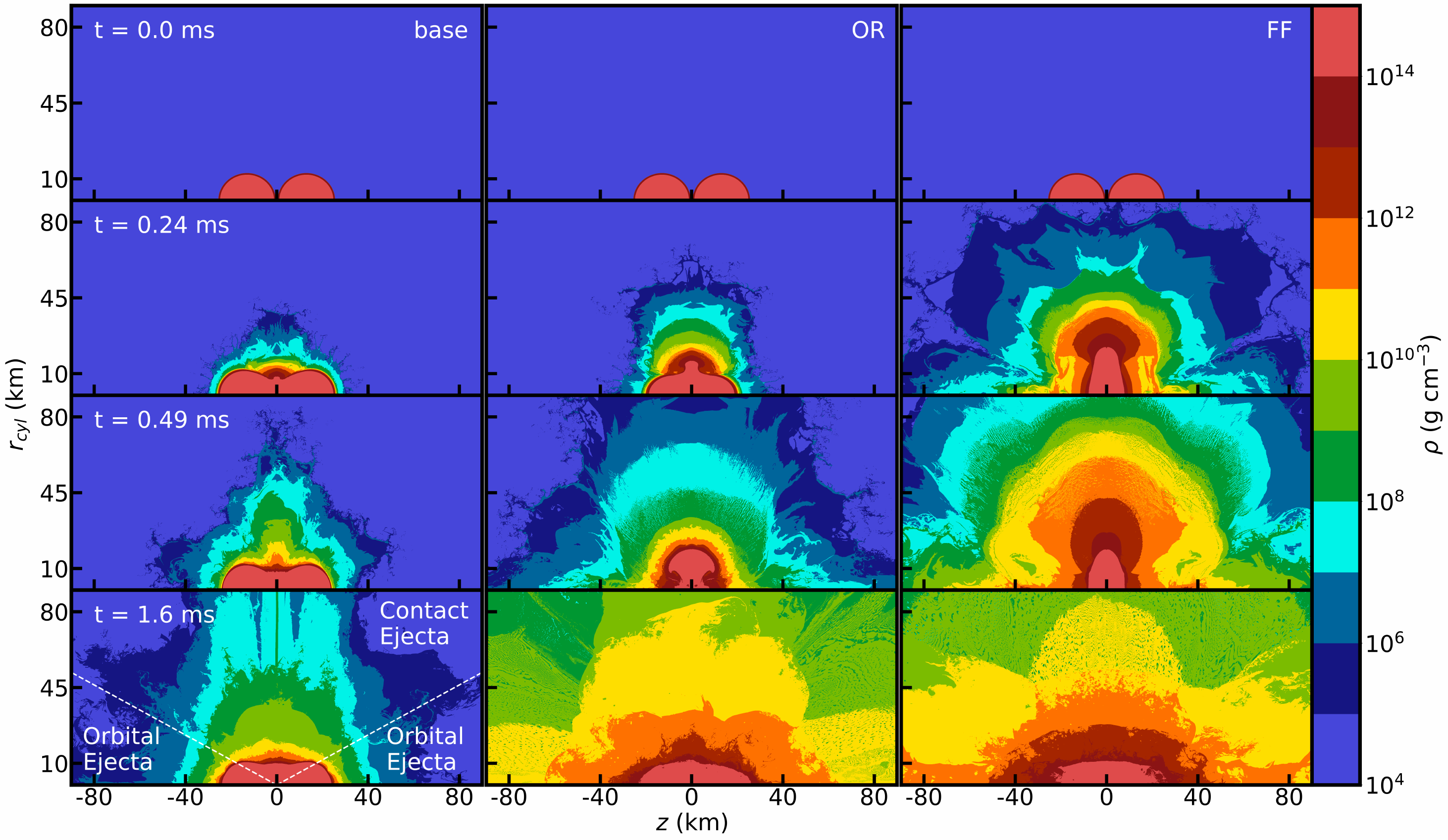}
\caption{\emph{Left:}
Snapshots in the evolution of the {\tt base} model (c.f. Table~\ref{tab:models}),
showing mass density at various times, as labeled. The white dotted line
at $30^\circ$ from the $z$-axis shows the division between \emph{contact} plane
and \emph{orbital} plane directions, as labeled.
\emph{Middle:} Same as in the left column but removing orbital motion, inertial forces, 
and gravitational wave losses (model {\tt OR}).
\emph{Right:} A head-on collision at the free-fall speed (model {\tt FF}), for comparison.
}
\label{fig:dens_parameterspace_timeseries}
\end{figure*}

\begin{figure*}
\centering
\includegraphics*[width=0.8\textwidth]{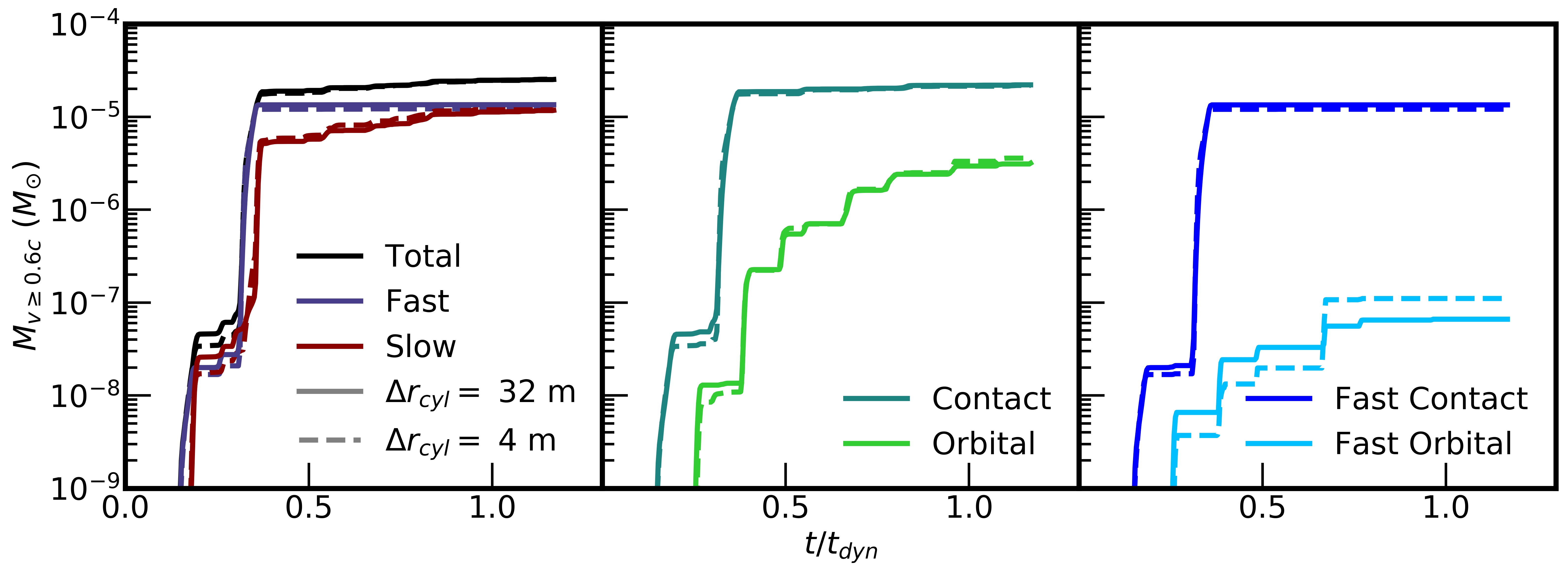}
\caption{Time-dependence of mass ejection in model {\tt base} (solid lines) and 
{\tt R4} (dashed lines, highest resolution), as measured from a spherical sampling surface 
with a radius $30$\,km from the origin.
\emph{Left:} Unbound ejecta separated into fast ($v>0.6$\,c) and slow ($v< 0.6$\,c).
\emph{Middle:} Total unbound ejecta separated into angular directions (c.f. Figure~\ref{fig:dens_parameterspace_timeseries}).
\emph{Right:} Fast ejecta separated into angular directions.
}
\label{fig:base_R4_temporal_evolution}
\end{figure*}

\subsection{Overview of baseline model and resolution dependence}
\label{sec:base_overview}

As the stars collide, mass is ejected on a timescale of $\sim 1$\,ms, first in the general direction of 
the rotation axis, and then toward equatorial regions, as 
shown in the left row of Figure~\ref{fig:dens_parameterspace_timeseries}. For analysis, we divide the ejection 
directions into \emph{contact} plane and \emph{equatorial} plane by a surface $30^\circ$ 
from the $z$-axis (the contact plane being the region that includes the rotation axis).
Ejected material is considered 
unbound from the system when it has positive Bernoulli parameter:
\begin{equation}
{\rm Be} = \frac{1}{2}\mathbf{v}^2 + \epsilon + \frac{P}{\rho} + \Phi > 0.
\end{equation}
We sample the unbound mass flux at a spherical extraction radius $r_{\rm out} = 30$\,km
from the origin\footnote{
{Changing the position of this extraction radius can change the
inferred fast ejecta mass by a factor $\sim 2$.}
}
(the $z$-axis is the symmetry axis for the mass ejection
measurement sphere). 
\emph{Fast} ejecta is defined as that with radial velocity $v > 0.6$\,c at $r=r_{\rm out}$,
while \emph{slow} ejecta is that with $v < 0.6$\,c.

Mass ejection is episodic (Figure~\ref{fig:base_R4_temporal_evolution}), 
with two initial bursts of mostly fast ejecta, 
up to a time of $\sim 0.4\,t_{\rm dyn}\simeq 0.5$\,ms.
Thereafter, slow ejecta continues to build up
beyond $\sim 1\,t_{\rm dyn}\simeq 1.4$\,ms. 
Each of these episodes is the result of oscillations in the collision 
remnant, as seen in global 3D merger simulations (e.g., \citealt{bauswein_2013}).
These oscillations show as steps in the cumulative ejected mass
as a function of time in Figure~\ref{fig:base_R4_temporal_evolution}.
While contact plane ejecta appears in only two bursts, with marginal
increases thereafter, orbital plane ejecta gradually builds up
through several oscillations, with the average velocity of
the ejected material decreasing with time.
The vast majority of the fast ejecta is launched toward the contact plane direction
in this model. 

Note that on the timescale of our simulation, {production of}
fast ejecta {is} largely {complete}, while the slow ejecta 
{is still increasing}
(Figure~\ref{fig:base_R4_temporal_evolution}). 
We thus find total
ejecta masses (fast and slow) that are significantly lower than those typically reported in global 3D merger 
simulations for this binary combination ($\gtrsim 10^{-3}M_\odot$; e.g., 
\citealt{hotokezaka_2013}, \citealt{rosswog_2013}, 
\citealt{lehner_2016}, \citealt{sekiguchi_2016}, \citealt{bovard_2017}, \citealt{dietrich_2017}, 
\citealt{radice_2018b})

\begin{figure*}
\centering
\includegraphics*[width=0.8\textwidth]{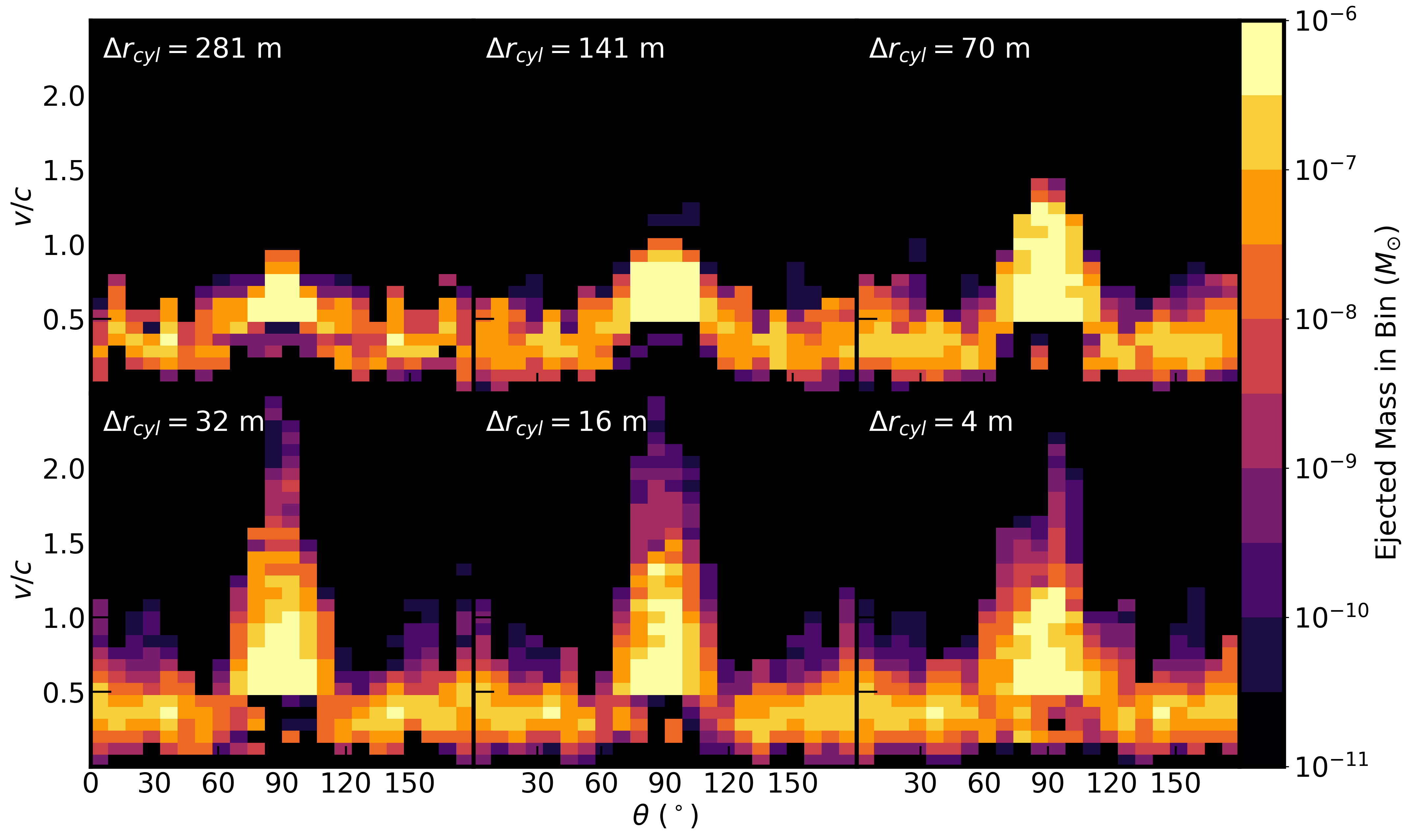}
\caption{Mass histogram of unbound ejecta as a function of ejection polar angle
(measured from the $z$-axis, c.f.~Figure~\ref{fig:dens_parameterspace_timeseries}, $90^\circ$
corresponds to the rotation axis) 
and velocity, for our default binary configuration. Different panels correspond to 
different spatial resolutions, as labeled (c.f. Table~\ref{tab:models}, the panel
for $\Delta r_{\rm cyl}=32$\,m corresponds to the {\tt base} model). 
Since our simulations are Newtonian, a small amount of mass
can achieve speeds larger than the speed of light.
}
\label{fig:resolution_2D}
\end{figure*}

Figure~\ref{fig:base_R4_temporal_evolution} also shows that the mass ejection
history is qualitatively the same in the baseline model and in the highest
resolution model (R4). In particular, the contact plane ejecta is the same
within $10\%$. While larger changes with resolution up to a factor $\sim 2$
are seen in the fast material ejected toward equatorial latitudes,
this contribution is subdominant compared to matter ejected toward
the contact plane.

The effect of spatial resolution on the angle- and velocity distribution
of cumulative unbound ejecta is shown in Figure~\ref{fig:resolution_2D}.
This sequence of models spans a factor of $\simeq 70$ 
in resolution,
from the coarsest grid size $\Delta r_{\rm cyl}\simeq 280$\,m to the finest
value at $4$\,m. While the overall shape of these two-dimensional
histograms is the same in all cases, the most visible change occurs for
matter within $\sim 30^\circ$ of the rotation axis, i.e. contact
plane ejecta. As the resolution is increased, more mass is ejected around
the rotation axis at the high- and low velocity ends. Note that
since our simulations are Newtonian, a small amount of mass achieves
$v>c$. 

\begin{figure}
\centering
\includegraphics*[width=1.0\columnwidth]{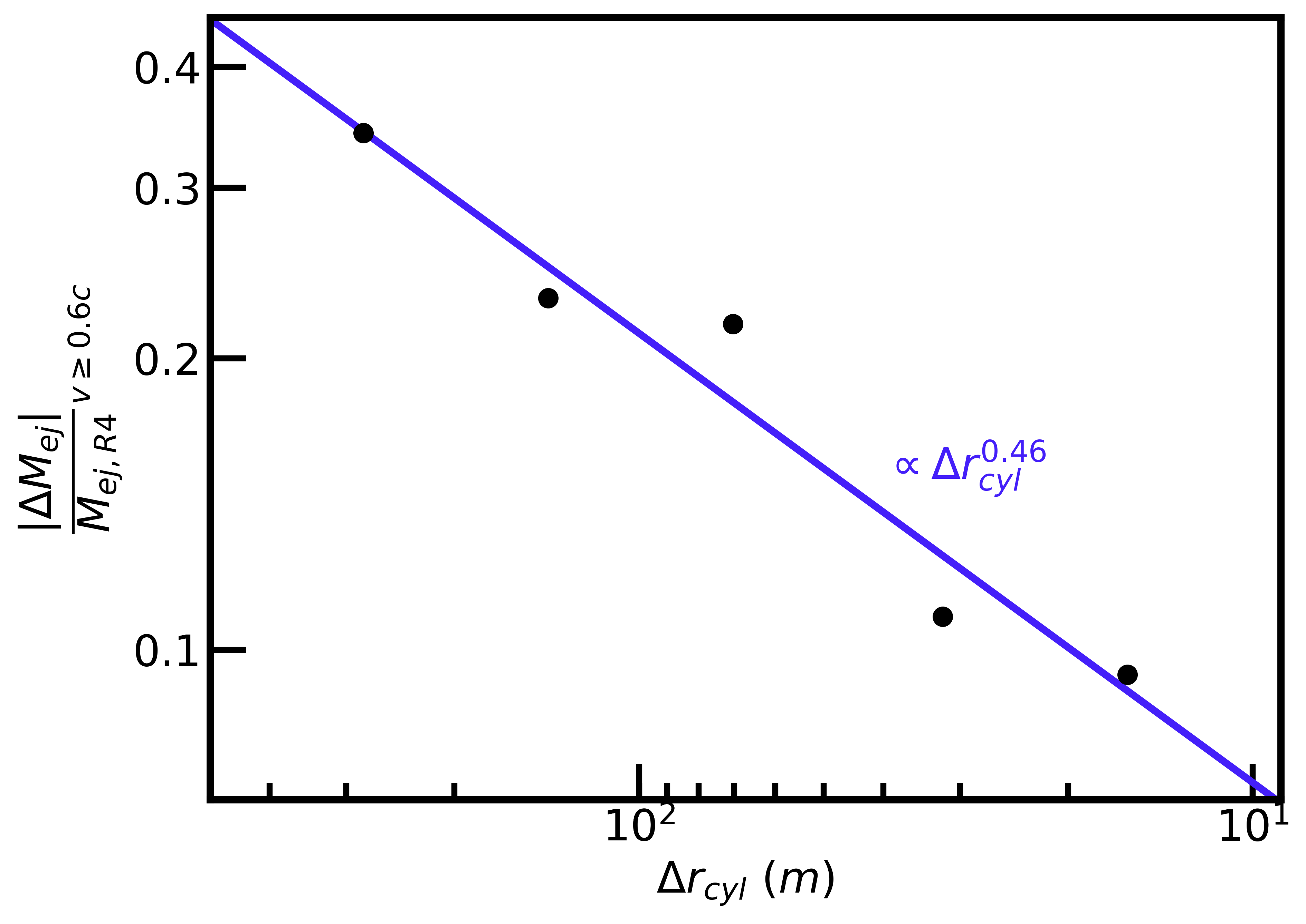}
\caption{{Convergence of fast ejecta mass with spatial resolution.
Shown are relative differences in total fast ejecta mass relative to that from
model R4. The purple line is a power-law fit.} 
}
\label{fig:convergence}
\end{figure}

{Taking the fast ejecta from model R4 as a baseline, we estimate the degree
of convergence of mass ejection by computing differences relative to this ejecta 
value as a function of resolution. Figure~\ref{fig:convergence} shows that the
total fast ejecta mass converges as $\Delta r_{cyl}^{0.46}$ over the entire
resolution range explored. Convergence to within $10\%$
is achieved for $\Delta r_{cyl}\approx19.7$\,m.}

A natural explanation of this trend with cell size is the increasing degree 
by which the stellar edges are resolved (Figure~\ref{fig:pressure_scale_height}).
Overall, fast ejecta on the contact plane can change by a factor of $\sim 2$
from the lowest to highest resolution (Table~\ref{tab:models}). The equatorial
ejecta also increases by a factor of $\sim 2$ over the entire resolution
range. Spatial resolution therefore also has an effect on matter ejected
during remnant oscillations (as inferred from Figure~\ref{fig:base_R4_temporal_evolution})
even though the overall shape of the equatorial ejecta distribution remains
largely the same in Figure~\ref{fig:resolution_2D}. Again, equatorial ejecta
contributes primarily with $v < 0.6$\,c in this model.

\begin{figure}
\includegraphics*[width=\columnwidth]{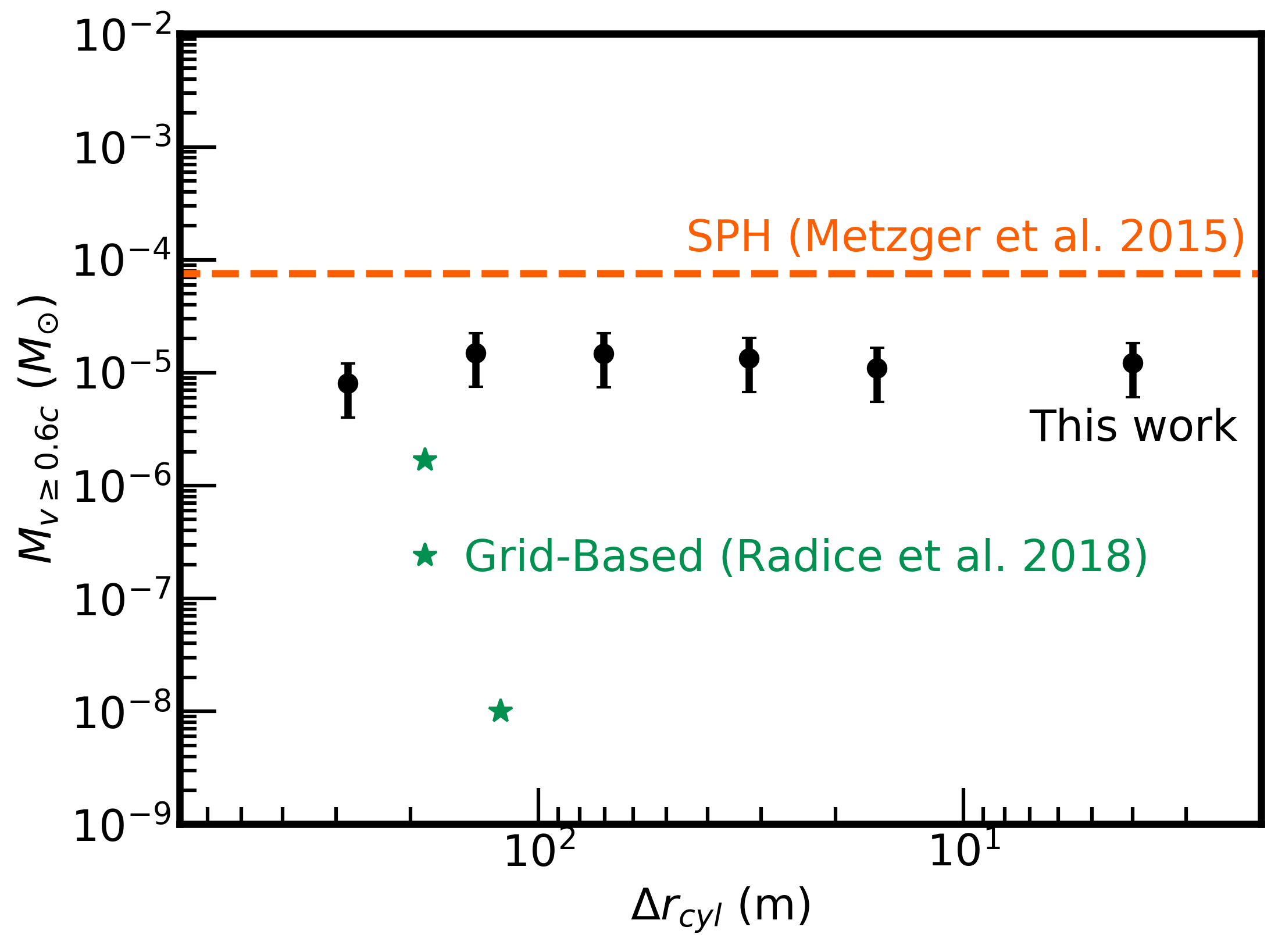}
\caption{Total fast ejecta for our default configuration (black circles) as a function 
of cell size (c.f. Table~\ref{tab:models}). 
The orange dashed line shows the 
$r$-process freezout ejecta
reported in \cite{metzger_2015} out of the equal-mass $1.35\Msun$ merger simulation
from \cite{bauswein_2013}, carried out with an SPH code without neutrinos 
and the \texttt{DD2} EOS.
Green stars show the fast ejecta reported in \citet{radice_2018b} for 
a set of models with the closest parameters to our {\tt base} case:
an equal-mass $1.35\Msun$ binary simulated with a grid code and the \texttt{DD2} EOS.
The resolutions and neutrino treatment are, from top to bottom, $185$\,m with an M0 scheme,
$185$\,m with a leakage scheme, 
and $123$\,m with a leakage scheme, respectively. {We estimate that our
values have a measurement uncertainty of a factor $\sim 2$ due to sensitivity to the extraction
radius and the imposition of a velocity cut at $0.6$\,c 
to estimate free neutron masses (\S\ref{sec:comparison}).}}
\label{fig:free_neutron_mass_resolution}
\end{figure}

The total fast ejecta from our default configuration across the resolution range
we explore is shown in Figure~\ref{fig:free_neutron_mass_resolution}, where
the very mild dependence with resolution is apparent. For comparison, we also
show fast ejecta results from the grid-based, global 3D merger simulations of
\citet{radice_2018b}, 
corresponding to equal-mass $1.35M_\odot$ binaries evolved with different resolutions and neutrino prescriptions.
Our results suggest that the fast ejecta values from \citet{radice_2018b} are already close
to convergence, and are unlikely to increase (for higher resolution) by the magnitude
required to match the amount inferred by \citet{metzger_2015} from the SPH simulations
of \citet{bauswein_2013}.

We also note that the initial burst in the {\tt base} model ejects
$\sim 10^{-8} - 10^{-7}M_\odot$ (Figure~\ref{fig:base_R4_temporal_evolution}), 
which has similar magnitude to the
ultrarelativistic envelope envisioned by \citet{beloborodov_2020} as
a breakout medium for the jet in order to account for the features
of the prompt emission in GRB170817A. We caution that for these small
amounts of mass, other processes such as neutrino emission and absorption
(as envisioned by \citet{beloborodov_2020} to power the expanding
envelope) or magnetic fields can be dynamically important and are missing
in our simulations. Also, the mass in ambient medium in our models
is $\sim 10^{-8}M_\odot$ (Section \ref{sec:initial_conditions}), hence
the first burst of ejecta is slowed down significantly and its final
velocity in our simulations is not physical. Nevertheless,
our results suggest the possibility that such a fast moving envelope
of small mass could arise from the hydrodynamic interaction alone.
Properly capturing this phenomenon will require highly-resolved global
simulations with neutrino transport and magnetic fields.

\vspace{0.2in}
\subsection{Dependence on force prescription and collision velocity}
\label{sec:force_prescription}

As an attempt to quantify the uncertainty in the approximations used in our
numerical experiment, we evolve a model ({\tt OR}) in which we remove the
corotation of the coordinate system: no orbital motion, and therefore no
centrifugal and Coriolis forces (Section \ref{sec:physical_model}), as well as no
gravitational wave losses (Section \ref{sec:hydrodynamics}). The setting is thus a
head-on collision at a speed set by the decay of the semimajor axis due to
gravitational waves (equation~\ref{eqn:adot_gw}). 

Figure~\ref{fig:dens_parameterspace_timeseries} compares the evolution
of model {\tt OR} with our {\tt base} case. Mass ejection occurs at a
faster pace, and spans a broader range of latitudes.
Ejecta is produced in an episodic manner, with two main bursts occurring 
during the simulated time, like in model {\tt base}. The second burst is more temporally spread than 
the first burst in the \texttt{base} case, however. Also, while in the latter model both bursts 
of fast ejecta are predominantly launched toward the contact plane, in model 
\texttt{OR} the first burst is launched toward the contact plane and the second toward
the orbital plane. We also note that 
model {\tt OR} produces more slow ejecta than fast ejecta.

Table~\ref{tab:models} shows that the total fast ejecta is a factor $\sim 300$ larger in model
{\tt OR} than in the {\tt base} case, with a comparable separation between
contact and equatorial plane directions. While the total ejecta mass shows
a non-monotonic dependence on spatial resolution, the fast component does
vary monotonically, with contact plane ejecta increasing
and orbital plane ejecta decreasing with finer grid spacing, and such that
the total fast ejecta decreases by $\sim 20\%$ when going from $\Delta z = 141$\,m
to $32$\,m. Lacking the suppressing effect
of centrifugal forces, we can take model {\tt OR} as an absolute upper
limit to the fast ejecta generated from a binary neutron star collision.

For reference, we also consider a head-on collision at the free-fall speed
(model {\tt FF}). Such a calculation has a long history, and the qualitative result is
well-known (e.g., \citealt{shapiro_1980,rasio_1992,centrella_1993,ruffert_1998,kellerman_2010}). 
This type of collision is astrophysically relevant in the context of eccentric mergers, in which
head-on or off-center encounters are a common outcome (e.g., \citealt{gold_2012,chaurasia_2018,papenfort_2018})

Figure~\ref{fig:dens_parameterspace_timeseries} shows that model {\tt FF} behaves
in a qualitatively similar way to model {\tt OR}, but due to the higher collision
speed, mass ejection occurs at a faster rate. Material first decelerates in the collisional
direction as the density gradients meet. The strong pressure gradients 
produced accelerate material in the contact plane, expanding in that direction 
before collapsing back and then expanding in the orbital plane
direction (e.g, \citealt{centrella_1993}). Over longer time periods, the remnant
oscillates in a pattern that alternates between the contact and orbital
planes, before settling into a spherical remnant.
Model \texttt{FF} exhibits two such oscillations of the collision remnant
during the $1.6$ ms evolution time, resulting in a series of
periodic fast ejecta bursts first toward the contact plane, then toward the orbital plane,
with the production of fast ejecta saturating before the end of the simulation. 
We do not follow the oscillation of the remnant for long enough to fully capture all slow
ejecta produced, however, which is typically launched by 4-6 violent oscillations of the remnant,
before settling into a spherical shape over a period of $\sim 3$ ms
\citep{ruffert_1998}.

Table~\ref{tab:models} shows
that mass ejection from model {\tt FF} is significant, with a total fast ejecta reaching $0.01M_\odot$.
Like in model {\tt OR}, the total unbound ejecta shows a non-monotonic dependence on resolution.
While the fast ejecta produced toward the contact plane decreases by $\sim 25\%$ for increasing
resolution, fast ejecta launched toward the orbital plane is dominant by a factor $\sim 50$ and 
shows non-monotonic behavior.

\subsection{Dependence on EOS, total mass, and mass ratio}
\label{sec:parameter_variation}

\begin{figure*}
\centering
\includegraphics*[width=0.8\textwidth]{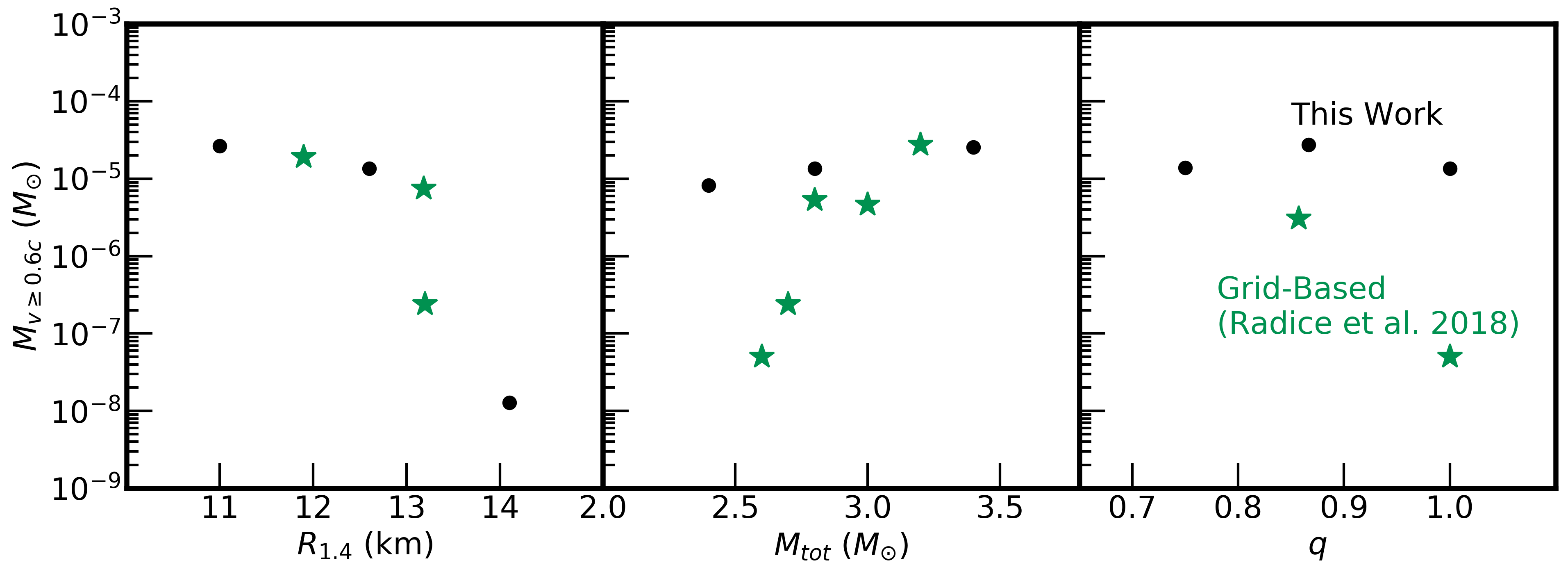}
\caption{Total fast ejecta from our simulations (black circles), for varying
EOS (left), total binary mass (center), and mass ratio (right). See Table~\ref{tab:models}
for model parameters.
For comparison, green stars show a subset of the fast ejecta results from 
\cite{radice_2018b} with the closest parameters.
Note that \cite{radice_2018b} also reports a merger with compactness
that falls between the leftmost two data points in the left panel, which 
produces no fast ejecta (\texttt{LS220\_M135135\_LK}). 
Similarily, 
two runs by \cite{radice_2018b} with total
masses of $2.4$ and $2.5 \Msun$ (\texttt{DD2\_M120120\_LK} and \texttt{DD2\_M125125\_LK}) also produce no fast ejecta. 
}
\label{fig:free_neutron_mass_parameter_space_alt}
\end{figure*}

When varying the EOS and therefore the NS radius relative to the baseline
configuration, we find a monotonically increasing dependence of the quantity of
fast ejecta produced on compactness
(Figure~\ref{fig:free_neutron_mass_parameter_space_alt}).  Our model with the
largest NS radius ({\tt BPAL12}) produces $\sim 1000$ times less fast ejecta
than our {\tt base} case, even though the total amount of ejecta is lower by a factor
of $20$ only: the vast majority of dynamical ejecta is slow. 
The oscillations of the remnant in this case are also much weaker and occur more frequent
than in the {\tt base} case, with the production of slow ejecta occuring over 4 or more 
bursts during the simulated time in contrast to the two bounces in the \texttt{base} case.
At the other end, our most compact configuration {\tt APR1} ejects only a factor $\sim 2$ more
mass than the {\tt base} model, both fast and slow.  More compact stars dive
deeper into the gravitational potential upon collision, and therefore more
energy becomes available to eject mass. This dependence of dynamical ejecta
quantity on compactness is a well-known result from global 3D merger
simulations (e.g., \citealt{hotokezaka_2013}). 

\begin{figure*}
\centering
\includegraphics*[width=0.85\textwidth]{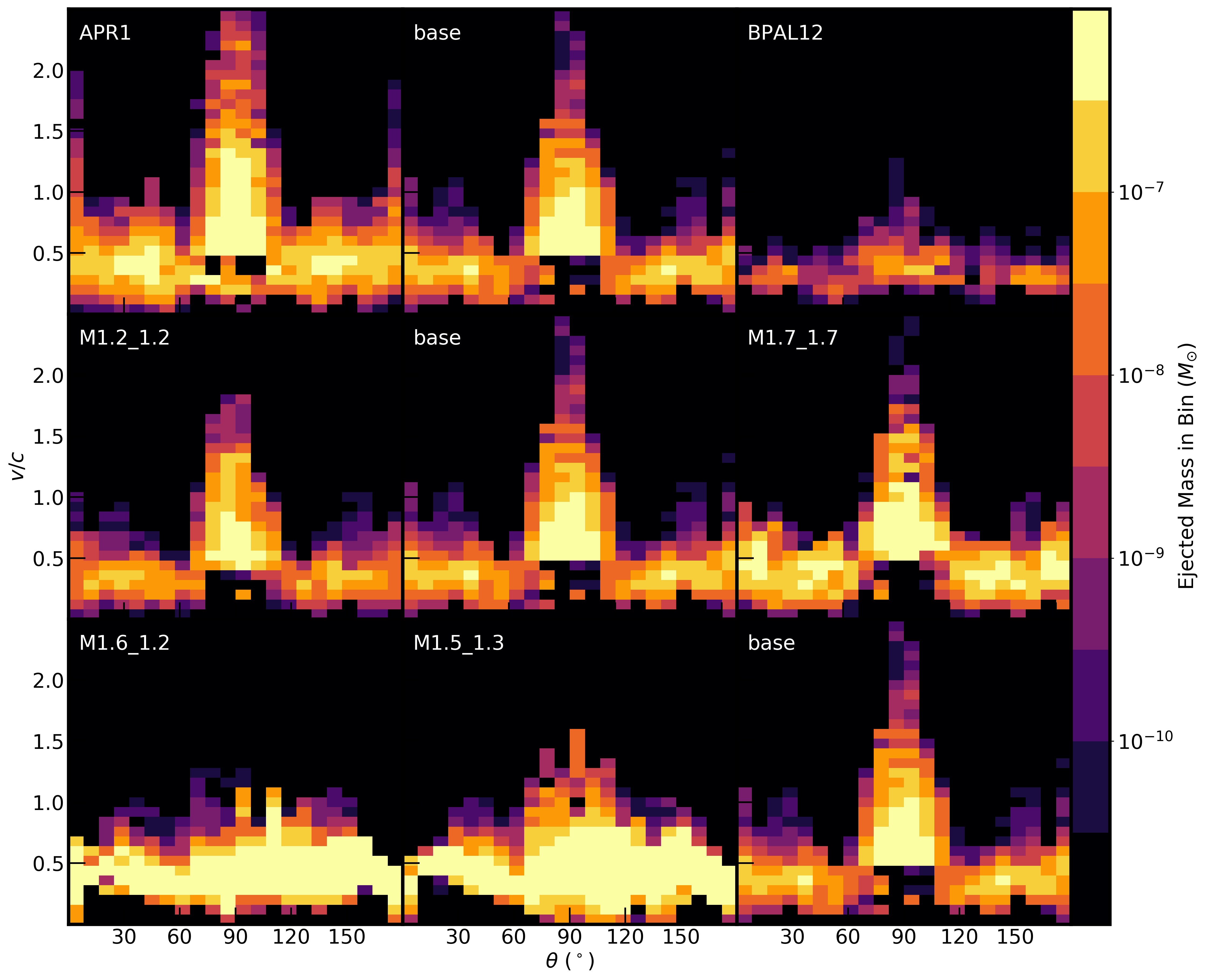}
\caption{
Same as Figure~\ref{fig:resolution_2D} but for models that vary the EOS (top row), total
binary mass (middle row), and mass ratio (bottom row), as labeled (c.f. Table~\ref{tab:models}).
}
\label{fig:parameter_space_2D}
\end{figure*}

The angle-velocity mass histogram (Figure~\ref{fig:parameter_space_2D}) shows
that for the least compact model {\tt BPAL12}, the high-velocity tail of the
contact plane ejecta disappears, yielding a more uniform distribution in
polar angle. The highest compactness model {\tt APR1} shows an extension of the
contact plane ejecta to even higher velocities, and an increase in the amount
of fast equatorial ejecta.

When varying the total mass of the binary system, we find a mild monotonic
dependence of fast ejecta with binary mass. Going from a $1.2+1.2M_\odot$
binary to $1.7+1.7M_\odot$, we find an increase by a factor $\sim 3$ of both
fast and slow ejecta masses
(Figure~\ref{fig:free_neutron_mass_parameter_space_alt}).
Our models, being Newtonian, exclude the possibility of prompt BH
formation, which is a likely outcome of the model with the highest 
total binary mass. Prompt BH formation is usually detrimental for mass
ejection (e.g., \citealt{hotokezaka_2013,bauswein_2013}).

Table~\ref{tab:models} shows that in all cases that vary the total mass, the fast orbital plane ejecta
remains subdominant compared to that launched toward the contact plane.  This
is consistent with the angle-velocity mass histogram
(Figure~\ref{fig:parameter_space_2D}), which shows very similar results for all
3 cases that change the total binary mass. For fixed EOS, more massive NSs
have smaller radii and therefore larger compactness, again yielding the known
trend of higher mass ejection for collisions that reach deeper into the
gravitational potential. 

Changing the mass ratio at fixed constant binary mass yields a non-monotonic
trend in fast ejecta production, with changes in total fast ejecta of a factor
$\sim 2$ (Figure~\ref{fig:free_neutron_mass_parameter_space_alt}).
By increasing the asymmetry of the binary, the number of ejection bursts increases 
from 2 bounces in the equal-mass \texttt{base} case to 4 or more oscillations for 
the asymmetric models (\texttt{M1.5\_1.3} and \texttt{M1.6\_1.2}). 
Additionally, orbital plane ejecta is predominantly launched (factor $\sim 3$) 
in the direction of the smaller star.

Table~\ref{tab:models} shows that the majority of fast ejecta continues to be
produced toward the contact plane, with overall changes in this subset
dominating the overall trend. An additional trend is a sharp increase in the
amount of equatorial fast ejecta with decreasing mass ratio by a factor $30$
over the range explored. However, even in the most asymmetric case
($1.6+1.2M_\odot$), the equatorial plane contribution to the fast ejecta is
lower by a factor $10$ than that from the contact plane.  The angle-velocity
histogram (Figure~\ref{fig:parameter_space_2D}) shows that increasing the
asymmetry of the binary results in a decrease of the high-velocity tail of the
contact plane ejecta, and a more isotropic production of ejecta. 

\subsection{{Discussion and} Comparison with previous work}
\label{sec:comparison}

Our baseline configuration yields fast ejecta quantities that fall in between
those found using SPH and grid-based methods, as shown in
Figure~\ref{fig:free_neutron_mass_resolution}. Given the number of
approximations necessary to make our numerical experiment possible, the
absolute amount of fast ejecta we find is not the most reliable quantity. Nonetheless, our
most important result is that there is little resolution dependence in this
fast ejecta when going from typical cell sizes employed in global 3D merger
simulations using grid-based codes ($\sim 100$\,m) to values that can capture
the dynamics at the stellar surface reliably (few m). 

We consider this
resolution dependence to be a robust result, 
{for the following reasons. First, the neutron stars
and post-collision remnant are spatially extended on scales $\sim R_{\rm NS}$.
The gravitational acceleration should be insensitive to small scale effects, except
if a strong curvature develops. The latter can arise with prompt black hole formation,
which is not considered in this study. Of particular concern would be high-mass
binaries very close to prompt collapse, in which a change in spatial
resolution could drastically change the amount of mass ejected. Second,
mass ejected at high speeds becomes length-contracted in the laboratory frame.
This effect
becomes significant for relativistic momenta $\gamma\beta\gtrsim 1$. 
In both our numerical experiment (Figure~\ref{fig:resolution_2D}) and in global 3D merger 
simulations, the ejecta has a mass distribution which is a decreasing function 
of velocity. We can estimate the error introduced by ignoring relativistic
kinematics by computing the ratio of fast ejecta with $v>c$ to that with $v>0.6c$ 
in the {\tt base} model: $M_{v\geq c}/M_{v\geq 0.6c} \simeq 0.17$. This $20\%$
uncertainty in our fast ejecta values due to the Newtonian description
is comparable to the changes with grid spacing in our best resolved models.
For comparison,
}
\cite{radice_2018b} finds variations of order a few in fast ejecta mass when
increasing the resolution from $185$\,m to $123$\,m. Our models yield a similar
resolution dependence when 
{changing the grid spacing} from $281$\,m to $141$\,m.

{The choice of extraction radius at $30$\,km does introduce a degree of uncertainty
in the absolute values of fast ejecta that we report. Increasing this extraction radius
to $90$\,km decreases the fast ejecta mass by a factor $\sim 2$ in the base model. Most of 
this fast ejecta leaves the computational domain before the end of the simulation, therefore this
dependence on the position of the extraction radius implies some degree of slowdown
in the vicinity of the collision remnant. Since all simulations are measured with the
same extraction radius, however, the position of this surface should not significantly affect
changes in fast ejecta with resolution.}

\begin{figure}
\includegraphics*[width=\columnwidth]{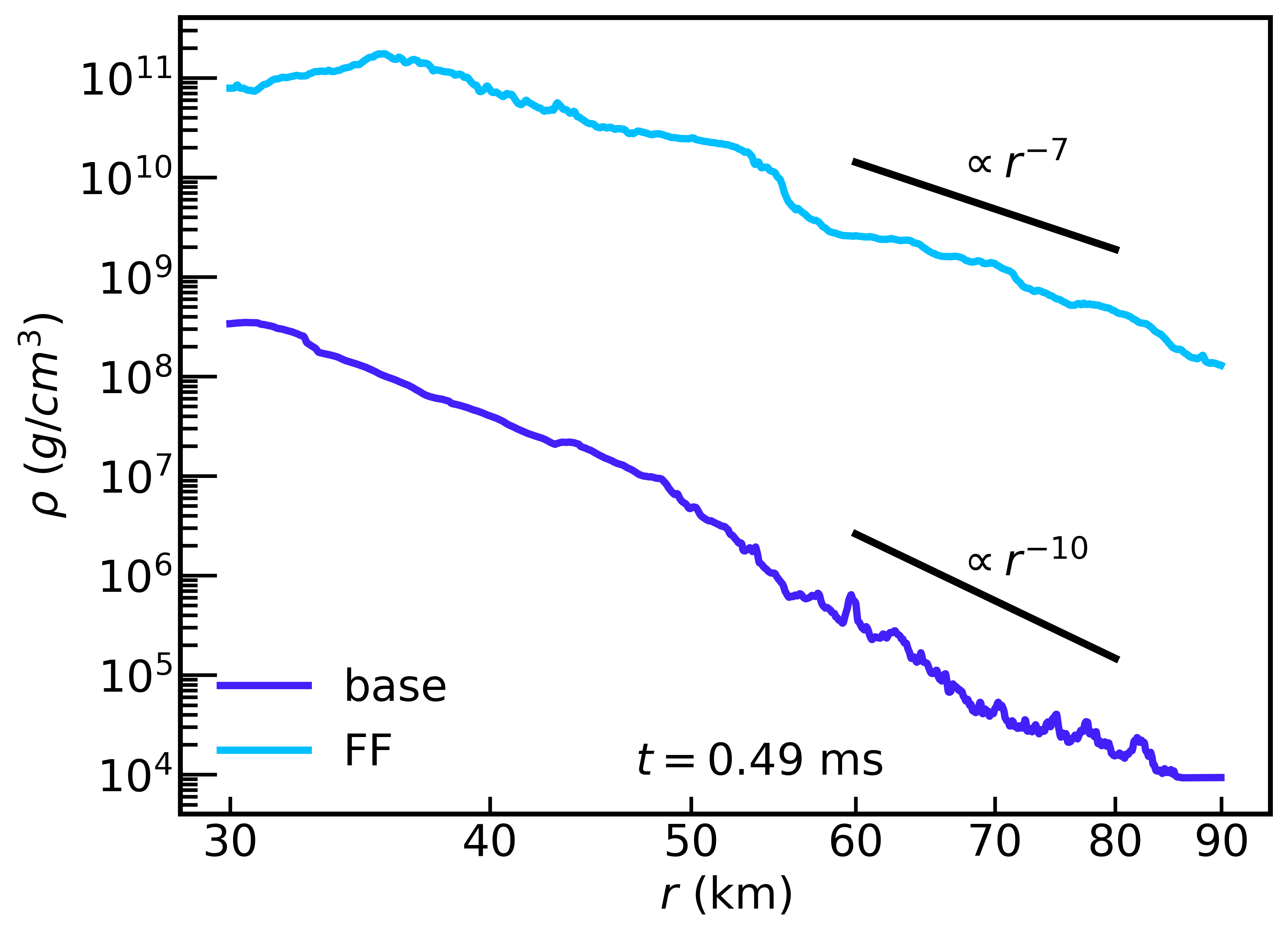}
\caption{Angle-averaged radial profiles of density at $0.49$\, ms
for models {\tt base} (purple) and {\tt FF} (blue), with the average carried out over the
contact and orbital plane directions, respectively.
Radial slopes of $r^{-7}$ and $r^{-10}$ are shown for reference.}
\label{fig:dens_r}
\end{figure}

Our definition of fast ejecta ($v>0.6$\,c) has been adopted to allow
direct comparison with previous work. For the purposes of estimating the
freezout of the $r$-process, however, a more strict criterion involves using the
expansion time, with the relevant material expanding to densities $4\times 10^5$\,g\,cm$^{-3}$
in less than $5$\,ms \citep{metzger_2015}. For completeness, we estimate
the expansion time of our fast ejecta material. In the absence of tracer
particles, we compute this estimate as follows. We first consider the angle-averaged
radial profile of the density of ejected material beyond the sampling radius
at a time $t=0.49$\,ms 
(Figure~\ref{fig:dens_r}), with the 
average carried out over the contact plane for the \texttt{base} model, and over the 
orbital plane for model \texttt{FF}, since these directions contain the 
majority of the fast ejecta (Table~\ref{tab:models}). Assuming the
material ejection is in steady state, we estimate an upper limit to the expansion 
time for the {\tt base} model as the crossing time from $30$\,km to $60$\,km at $v=0.6$\,c,
at which point the target density for freezout is reached. This yields an expansion 
time of $1.6\times 10^{-4}$\,s. For the \texttt{FF} case, we extrapolate the density profile
and find the distance necessary to reach the target density for freezout,
and compute the crossing time at $v=0.6c$. The result is 
a slightly longer $5.8\times 10^{-4}$\,s. In both cases, these simple
estimates indicate that our velocity cut at $0.6$\,c is more strict than what
is needed to achieve neutron freezeout, with a corresponding underestimate 
of the mass available to power a precursor. Lowering the velocity cut in our 
numerical experiment would increase our ``fast" ejecta at most by a factor of $\sim 2$, however,
since our slow ejecta is of comparable magnitude (Table~\ref{tab:models}). But
since global 3D merger simulations produce significantly more slow ejecta, the exact 
value of the velocity cut can be in part responsible for the large discrepancy
with particle-based results. The velocity distribution of ejecta in Figure 1 of \citet{metzger_2015}
shows that there is indeed non-negligible free neutron production for $0.4 < v/c < 0.6$.
A more careful analysis of this question requires use of 
sufficient tracer particles to sample the fraction of the velocity distribution that will 
yield the ejecta of interest.

{Given the extraction radius dependence of our fast ejecta values, and the possibility
that more free neutrons can be produced at $v<0.6c$, we can adopt a fiducial uncertainty
of a factor $\sim 2$ in the absolute values of our fast ejecta.}

Since most of our simulations are only run long enough for production of fast ejecta to be
complete, 
slow ejecta is still being produced by
the end of the simulation. 
By the end of our {\tt base} model, 
$2.5 \times 10^{-5}\Msun$ of total ejecta is produced. 
This is roughly an order of magnitude less ejecta than the closest 
case in
\citet{radice_2018b}, and about two orders of magnitude less
ejecta than was produced in the SPH simulation performed by \cite{bauswein_2013}.
While short run times (set by computational resources) are partly
responsible for this lack of slow ejecta, our lower resolution models
({\tt R281, R141, R70}) which run for ten times longer than the {\tt base} case
still do not produce sufficient slow ejecta to match 3D global
simulations. For example, model \texttt{R70} produces $1\%$ more fast ejecta
and a factor of $\sim 2$ more slow ejecta for an increased runtime a factor of 10 longer.
The imposition of axisymmetry and other approximations therefore
also play a role in this discrepancy.

An important difference between our results and those of \citet{radice_2018b}
is the way in which most of the fast ejecta is produced, a likely
consequence of the approximations made in our study.
\cite{radice_2018b} finds that fast ejecta emerges in two bursts, first from the
contact interface and then from the first bounce of the HMNS remnant. The first burst
is not always present, but the HMNS bounce is more robust and dominates
the production of fast ejecta. 
We see a two burst structure in many of our simulations when varying total mass,
or EOS (except in the low compactness case \texttt{BPAL12}, which produces less contact plane ejecta). 
This is also apparent in our force prescription variations, with \texttt{FF} producing
an additional late time burst. This burst structure however is largely a
temporal variation, with the majority of the ejecta in each burst being launched
in the contact plane, whereas \cite{radice_2018b} finds the first burst to be
launched toward the contact plane, and the second burst being distinctly constrained to
the orbital plane. 

The direction of the trends of fast ejecta quantity with EOS and total mass
in our models (Figure \ref{fig:free_neutron_mass_parameter_space_alt}) compare
favorably with those found by \cite{radice_2018b}, with our dependence
on total mass being milder.
Stronger differences are found when considering the mass ratio:
\cite{radice_2018b} obtains a strong monotonic dependence, whereas our
models yield a non-monotonic trend with changes of a factor $\sim 2$ only. Changing the
mass ratio does however result in a more broad angular distribution in our
simulations, with a larger fraction of fast ejecta being launched in the orbital 
plane in more asymmetric binaries.

\section{Implications for Kilonovae}
\label{sec:em_implications}

\subsection{Neutron-powered precursors}

For a given quantity of forward fast
ejecta $M_{\rm n}$ that freezes out the $r$-process, which here
we take to be fast ejecta with $v\ge 0.6$\,c, the peak bolometric luminosity of the ultraviolet
precursor is given by
\citep{metzger_2015}
\begin{equation}
\label{eq:lum_n_peak}
L_{\rm peak} = M_{\rm n} \dot{e}_n \lesssim 5\times10^{42}\left(\frac{M_{\rm n}}{10^{-5}\,\Msun}\right)\textrm{\,erg\,s}^{-1} 
\end{equation} 
where 
\begin{equation}
\dot{e}_n = 3.2 \times 10^{14} (1-2Y_e) \textrm{\,erg\,s}^{-1}\,\textrm{g}^{-1}.
\end{equation}
is the specific heating rate due to free neutron decay, and we have used
an electron fraction $Y_e = 0.1$ to estimate an upper limit for 
the free neutron mass fraction in evaluating equation~(\ref{eq:lum_n_peak}).
The corresponding time to peak emission is
\begin{equation}
\label{eq:time_n_peak}
t_{\rm peak} \simeq 0.3\,{\rm h\,}\left(\frac{M_{\rm n}}{10^{-5}\,\Msun}\right)^{1/2} 
                           \left( \frac{\kappa}{1\,\textrm{cm}^2\,\textrm{g}^{-1}}\right)^{1/2}
                           \left(\frac{v}{0.6\,\textrm{c}}\right)^{-1/2},
\end{equation}
where $\kappa$ is the effective gray opacity of the ejecta, the value of which can
vary in the range $1-30$\,cm$^{2}$\,g$^{-1}$ depending on the composition
and ionization state of the ejecta (e.g., \citealt{tanaka_2018}).
The fast ejecta produced in our baseline configuration, $M_{\rm n}\sim 10^{-5}\,M_\odot$ almost
independent of spatial resolution, would
correspond to the estimates above. 
Extrapolating the light curve models of \citet{metzger_2015} to lower neutron
masses, one would predict that for $\sim 10^{-5}\,M_\odot$ of free neutrons, the precursor
emission would peak around U-band (365 nm) at an AB magnitude $\approx 23$ for
a source distance of $200$\,Mpc.  The planned wide-field
ULTRASAT\footnote{\url{https://www.weizmann.ac.il/ultrasat/}} satellite mission
\citep{sagiv_2014} is expected to reach a 5$\sigma$ limiting magnitude of $\sim
23$ for a 1-hour integration at a wavelength of 220-280\,nm.  ULTRASAT
observations of nearby mergers (< $200$\,Mpc) could in principle probe the
neutron precursor emission given this level of fast ejecta production.

When we consider our main result of a weak dependence of $M_{\rm n}$ on spatial resolution for 
grid-based simulations, and the normalization of this quantity obtained in the global 3D
merger
simulations of \citet{radice_2018b} (e.g., Figure~\ref{fig:free_neutron_mass_parameter_space_alt}),
we conclude that the neutron-powered precursor luminosities and
timescales in equations~(\ref{eq:lum_n_peak})-(\ref{eq:time_n_peak}) are likely
to be an upper limit to luminosity and duration of this transient, at least as arising from the 
prompt dynamical ejecta. Other ways to produce fast ejecta include 
magnetically-accelerated, neutrino-heated winds from the magnetized neutron star remnant \citep{metzger_2018,ciolfi_2020},
outflows from accretion disks with strong initial poloidal fields \citep{fernandez_2019}, 
jet-cocoon interactions \citep{gottlieb_2020},
or ablation from an early neutrino burst generated from the colliding stellar interface \citep{beloborodov_2020}.
More reliable estimates of the fast ejecta from the entire merger event will therefore require
realistic initial conditions and a complete physics description, including MHD and neutrino transport
effects.

The free-fall, head-on collision case yields $\sim 10^{-2}M_\odot$ of fast ejecta, 
with $\sim 10\%$ variation with spatial resolution. Once the time to peak becomes
longer than the free neutron decay time, the luminosity estimate in equation~(\ref{eq:lum_n_peak})
is no longer valid. Nevertheless, we expect this fact ejecta to contribute to the
early kilonova signature, which should contain the imprint of much faster expansion velocities
than those inferred from GW170817 and therefore provide a distinct signature of
eccentric mergers.

\subsection{Non-thermal afterglows}

The dynamical ejecta is predicted to generate its own non-thermal afterglow after
expanding sufficiently to interact with the interstellar medium on a timescale
of years \citep{nakar_2011}. The duration and brightness of the afterglow
depend primarily on the kinetic energy distribution of the ejecta as well as
on the interstellar medium density (e.g., \citealt{kathirgamaraju_2019}).
Fast dynamical ejecta can produce an afterglow that evolves on
shorter timescales of $\sim $ months \citep{hotokezaka_2018}, and such a fast
component has been proposed as a possible origin of the X-ray excess recently
detected in the non-thermal GW170817 emission \citep{hajela_2021,nedora_2021}.

Here we focus on the implications of our resolution study for the
expected kilonova afterglows from BNSs. Given that our simulations are Newtonian,
we cannot produce a reliable relativistic momentum distribution. Nevertheless,
we can compute the kinetic energy of all the ejecta above a given velocity,
and look at its resolution dependence.

Table~\ref{tab:models} shows the kinetic energy of all the ejecta with radial velocities
greater than $0.3$\, and $0.6$\,c. The first velocity was found by \citet{kathirgamaraju_2019}
to be the minimum value needed to obtain a detectable afterglow. Our
baseline configuration yields characteristic kinetic energies $\sim 10^{49}$\,erg,
a factor $100$ below what is needed to generate a detectable afterglow. This is
likely a consequence of the underproduction of slow ejecta by our numerical 
experiment relative to global 3D merger simulations (\S\ref{sec:comparison}), given that most of the
kinetic energy normally resides in slower ejecta.

Despite this underproduction, we find that there is a convergence in the kinetic
energy (above both velocities) {to within $10\%$} for a resolution of $16$\,m, 
similar to the convergence in mass.
This is consistent
with the estimates of \citet{kyutoku_2014} for the characteristic cell size
at which shock breakout is properly resolved. 
{We note however that, in contrast to the mass, relativistic
material can carry a significant fraction of the total energy, hence the
lack of relativistic kinematics makes our estimates of energy dependence
on resolution only suggestive.}
{Our} results {nevertheless indicate} that 
robust predictions about the kilonova afterglow from the dynamical ejecta
require higher spatial resolution than used to date in global 3D merger simulations.

\section{Summary}
\label{sec:summary}

We have carried out a numerical experiment to probe the sensitivity to spatial
resolution of fast ejecta ($v>0.6$\,c) generation in grid-code simulations of
binary neutron star mergers.  Discrepancies in the amount of this ejecta of an
order of magnitude or larger have been found in global 3D merger simulations
using grid codes compared to the amounts found in SPH simulations. 
We implement an axisymmetric model of stellar collision in a co-rotating frame,
including the effects of inertial forces and gravitational wave losses
(Section \ref{sec:physical_model}-\ref{sec:hydrodynamics}).  The lower
computational expense of this setup allow us to probe spatial resolutions up to
$4$\,m, or $\sim 3\times 10^{-4}$ of the neutron star radius, smaller than the finest grids
used thus far in global 3D simulations of neutron star mergers by a factor of
$\sim 20$, and capturing the surface pressure scale height of the stars above
within $0.1\%$ of the stellar surface
(Figure~\ref{fig:pressure_scale_height}).  Our main results are the following:
\newline

\noindent 
1. -- Our baseline configuration of two $1.4M_\odot$ NSs with radius $12.6$\,km
ejects $\sim 10^{-5}M_\odot$ of fast ejecta, with variations in this quantity
of at most a factor of $\sim 2$ over a factor $140$ in cell size, converging 
{to within $10\%$} at a spatial resolution of {$20$\,m} 
(Figure~\ref{fig:free_neutron_mass_resolution}). 
While the absolute amount of ejecta is influenced by the approximations made in our 
reduced dimensionality experiment, the sensitivity to spatial resolution should be a robust outcome
{because of lack of strong curvature effects and a limited fraction of the ejecta
moving at relativistic speeds that introduce significant length-contraction effects}.
We conclude that existing grid-based, global 3D hydrodynamic simulations of binary
NS mergers (e.g., \citealt{radice_2018b}) are close to converging in the amount of 
fast ejecta, and the known discrepancy with SPH simulations is unlikely to be
caused by lack in resolution. A quantity of fast of ejecta of the order 
of $\sim 10^{-5}M_\odot$ is detectable with upcoming space-based UV facilities 
out to $\lesssim 200$\,Mpc
(Section \ref{sec:em_implications}).
\newline

\noindent
2. -- A head-on collision at the free-fall speed can eject $\sim 10^{-2}M_\odot$ of
both fast and slow ejecta (Figure~\ref{fig:dens_parameterspace_timeseries},
Table~\ref{tab:models}).  For the case in which such a collision becomes
possible due to a highly eccentric merger, the corresponding kilonova
signature can be distinct from that arising in a more conventional
quasi-circular merger driven by GW emission. When varying the resolution by a
factor of $4$, we find that the total ejecta from our head-on collision models 
can vary non-monotonically at the $\sim 20\%$ level. 
\newline

\noindent
3. -- Our numerical experiment reproduces the overall trends of fast ejecta
generation with compactness (EOS) and total binary mass 
(Figure~\ref{fig:free_neutron_mass_parameter_space_alt}) when compared to the 
global 3D study of \citet{radice_2018b}. We find a non-monotonic dependence
on mass ratio, which differs from that found previously. These differences
can be attributed primarily to the approximations made in our experiment,
particularly the reduced dimensionality and functional form of gravitational
wave losses (Section \ref{sec:physical_model}-\ref{sec:hydrodynamics}). While
we observe production of fast ejecta in bursts in our default configuration
(Figure~\ref{fig:base_R4_temporal_evolution}), the nature of these bursts is 
such that ejection is primarily in the contact plane direction, in contrast
to the global simulations of \citet{radice_2018b} in which the second, more
robust burst is confined to the equatorial plane. Additionally, our models preclude
the possibility of prompt BH formation, which can occur at high total binary
masses and result in reduced ejecta production relative to what we find.
\newline

\noindent
4. -- The kinetic energy of the fast ejecta {has a
similar dependece on} resolution
than the mass. 
This suggests that predictions for the non-thermal kilonova afterglow
require global 3D simulations with higher resolutions than those done thus far.
\newline

{Four} additional channels {have been proposed} for the production of fast ejecta: 
neutrino-driven outflows from magnetized neutron star remnants \citep{metzger_2018, ciolfi_2020},
outflows from accretion disks with strong initial poloidal fields \citep{fernandez_2019},
cocoon-jet interactions \citep{gottlieb_2020}, or ablation of stellar material
by neutrinos \citep{beloborodov_2020}. A more realistic estimate
of the total amount of fast ejecta produced in a binary neutron star merger
thus requires (1) modeling the collision in general-relativistic MHD with adequate 
neutrino transport and with sufficient spatial resolution, and (2)
use of realistic post-merger initial conditions in the case of fixed-metric 
codes are used. Such a calculation would provide useful information beyond
the fast ejecta, and therefore is a valuable direction to pursue.

The head-on collision of neutron stars at the free-fall speed remains a simpler
problem which, when done in numerical relativity and with sufficient
spatial resolution,
can yield useful predictions for the ejecta components from eccentric mergers,
allowing an estimate of the electromagnetic signal and nucleosynthesis contributions
from these collisions.

\acknowledgments
We thank Steven Fahlman, Mario Ivanov, Sharon Morsink, and Erik Rosolowsky 
for helpful discussions.
{We also thank the anonymous referee for helpful comments that improved the paper.}
CD and RF acknowledge support from the Natural Sciences and Engineering Research
Council of Canada (NSERC) through Discovery Grant RGPIN-2017-04286, and from the
Faculty of Science at the University of Alberta. 
CD also acknowledges support from the Alberta Government via the Queen Elizabeth II
Scholarship and Alberta Graduate Excellence Scholarship.
BDM acknowledges support from NASA (Swift Guest Investigator Program grant 80NSSC20K0909) 
and the National Science Foundation (grant AST-2002577). 
The software used in this work
was in part developed by the U.S. Department of Energy (DOE) 
NNSA-ASC OASCR Flash Center at the University of Chicago. 
Data visualization was done in part using {\tt VisIt} \citep{VisIt}, which is supported
by DOE with funding from the Advanced Simulation and Computing Program
and the Scientific Discovery through Advanced Computing Program.
This research was enabled in part by support provided by WestGrid
(www.westgrid.ca), the Shared Hierarchical Academic Research Computing Network
(SHARCNET, www.sharcnet.ca), Calcul Qu\'ebec (www.calculquebec.ca), and Compute
Canada (www.computecanada.ca).
Computations were performed on the \emph{Niagara} supercomputer at the SciNet 
HPC Consortium \citep{loken_2010,ponce_2019}. SciNet is funded by the Canada 
Foundation for Innovation, the Government of Ontario (Ontario Research Fund - Research Excellence), 
and by the University of Toronto. 

\software{
          {\tt FLASH} version 4.5 \citep{fryxell00,dubey2009}
          {\tt matplotlib} \citep{hunter2007}
          {\tt NumPy} \citep{harris2020array},
          {\tt VisIt} \citep{VisIt}
}


\vspace{0.1in}
\bibliographystyle{aasjournal}
\bibliography{ms}

\label{lastpage}

\end{document}